\documentclass[preprint2]{aastex631}
\DeclareUnicodeCharacter{2060}{}
\usepackage{comment}
\usepackage{lmodern}
\usepackage{anyfontsize}
\usepackage{hyperref}
\usepackage{amsmath}

\begin{document}

\title[Short title, max. 45 characters]{Observational biases on rotation curves from IFU data at cosmic noon}

\correspondingauthor{Amanda E. de Araujo-Carvalho}
\email{amanda@ov.ufrj.br}

\author[0000-0001-5539-0008]{Amanda E. de Araujo-Carvalho}
\author[0000-0003-2374-366X]{Thiago S. Gonçalves}
\affiliation{Valongo Observatory, Federal University of Rio de Janeiro, Ladeira do Pedro Antônio 43, Rio de Janeiro, RJ 20080-090, Brazil}

\author{Davor Krajnović}
\affiliation{ Leibniz-Institut für Astrophysik Potsdam (AIP), An der Sternwarte 16, 14482 Potsdam, Germany}

\author[0000-0003-3153-5123]{Karín Menéndez-Delmestre}
\affiliation{Valongo Observatory, Federal University of Rio de Janeiro, Ladeira do Pedro Antônio 43, Rio de Janeiro, RJ 20080-090, Brazil}

\author[0009-0005-1424-6604]{Natanael de Isídio}
\affiliation{European Southern Observatory, Karl-Schwazschild-Straße 2, 85748 Garching bei München, Germany }

\begin{abstract}
Through studying rotation curves, which depict how the velocity of the stars and gas changes with distance from the center of the galaxy, it has been confirmed that dark matter dominates galaxy's outer regions, as their rotation curve remains flat. However, recent studies of star-forming galaxies at cosmic noon have shown a decline in their rotation curve beyond a certain point, suggesting a decrease of the abundance of dark matter in galactic halos during earlier times. In this work, we investigate the influence of cosmological surface brightness dimming and loss of resolution on observations of rotation curves at cosmic noon. We used a sample of 19 Lyman Break Analogs at z $\sim$ 0.2 and artificially-redshifted them as if they were at z $\sim$ 2.2. By comparing both rotation curves of the observed and mocked objects, we find that the asymmetry of the cosmic noon galaxies is smaller than that of the low-z galaxies. In low-z galaxies, asymmetry increases with radius and becomes relevant at the external parts, where mergers and interactions causes more disturbance in galaxy's gravitational field. In contrast, cosmic-noon galaxies appear smoother, smaller and suitable for dynamical modeling - when in reality, they are not. The combined effects of the cosmological bias and loss of resolution lead us to the conclusion that caution should be exercised when using cosmic-noon rotation curves, as they might not accurately trace the gravitational potential of the galaxy.

\end{abstract}

\section{Introduction}
\label{sec:intro}

Currently, the most accepted hypothesis for how structures were formed in the early universe is the hierarchical $\Lambda$ cold dark matter ($\Lambda$CDM) model. According to this theory, initial small density perturbations collapsed, leading to the formation of dark matter halos \citep[e.g.,][]{zavala2019dark}. These halos were amplified by gravity, generating sufficient potential to concentrate baryons \citep{peebles1969origin}. As soon as the baryons began to cool, they collapsed into a disk (due to angular momentum), cooled further forming stars, and resulting in a rotating stellar disk \citep{1980MNRAS.193..189F,1984ApJ...286...38W}. The disk then evolves mainly through mergers with other systems and accretion of the surrounding cool gas, which later transforms into stars \citep{tiley2020galaxy}.

Due to observational challenges to verify this theory, cosmological simulations are often employed, providing credible predictions for the effects of galaxy formation processes on dark matter distribution. It is important to recognize, however, that simulations, such as IllustrisTNG \citep{pillepich2018simulating, pillepich2019first,nelson2019illustristng}, still faces notable challenges in accurately reproducing fundamental quantities such as the stellar and dark matter content of galaxies \citep[e.g.,][]{marasco2020massive, mancera2025galaxy}. These limitations underscore the need for continued improvements in the modeling of baryonic physics, especially the complexity between accretion and feedback processes, in order to fully understand galaxy dynamics on small scales. Although such constraints exist, the TNG simulations has successfully reproduced many observed galaxy properties across cosmic time \citep{vogelsberger2014introducing, snyder2015galaxy}.Among them, we can cite the large scale structures of the Universe and their evolution with cosmic time \citep{2014Natur.509..177V}, 
as well as the internal velocity structure of massive ($M_* \geq 10^{10-11} M_\odot$) late-type galaxies \citep{10.1093/mnras/stu1536}. This last work predicted that the inner parts of the galaxies at z = 0 are clearly dominated by the stellar mass. Beyond twice the stellar half-mass radius, the dark matter contribution to the circular velocity curve dominates over the stellar contribution for all galaxies. The combined stellar and dark matter circular velocity curves results in a nearly flat velocity profile extending to larger radii.

The flat rotation curve behavior is consistent with observations of galaxies' rotation curves in the local universe, providing some of the most compelling evidence for the existence of dark matter \citep[e.g.,][]{rubin1970rotation,1981AJ.....86.1791B,1981AJ.....86.1825B,2008AJ....136.2648D, posti2019peak,ren2019reconciling,li2020comprehensive,pina2021baryonic,pina2021tight,mancera2022impact,di2021rotation,di2023dark,de2024dark}. 
According to the \citet{2016A&A...594A..13P}, dark matter constitutes approximately 85\% of the universe's total mass. Despite its gravitational influence, the fundamental nature of dark matter remains elusive, with many open questions about its properties and evolution over time. 
Dynamical studies of more distant galaxies thus become essential for furthering our understanding of dark matter. 

The dynamics of stars and gas within galaxies is mainly influenced by the gravitational potential, determined by the mass distribution of dark matter and baryons. As galaxies evolve, the signature of their evolution remains in their kinematics \citep{2016MNRAS.461..859S, osman2017strong, pina2021baryonic, de2024dark}. In this context, recent studies have utilized different dynamical tracers to probe some of the dark matter properties \citep{10.1093/mnras/sty3404, 10.1093/mnras/stw3101, chakrabarti2021measurement,de2024dark}. The development of integral field units (IFUs) has allowed mapping of galaxy dynamics and emphasized the wide range of complex kinematics features that they can display (SINFONI \citep{2006ApJ...645.1062F,2009ApJ...706.1364F, Law_2009}, SAURON \citep{bacon2001sauron, tim2002sauron},MANGA \citep{2015ApJ...798....7B}, CALIFA \citep{10.1093/mnras/stx2409},SAMI \citep{10.1093/mnras/stv2643}, ATLAS$^{3D}$ \citep{2011MNRAS.413..813C}), KMOS$^{3D}$\citep{wisnioski2015kmos3d}, MUSE \citep{10.1093/mnras/stx201,2019MNRAS.485..934T}.

In particular, kinematic studies of the warm ionized gas have revealed low gas turbulence and a rotation-to-dispersion velocity ratio ($V/\sigma$) of approximately 1 in star-forming galaxies at cosmic noon. On the other hand, their low-redshift counterparts exhibits ratios of V/$\sigma \sim 10-20$ and higher gas turbulence in their disks \citep{Forster,wisnioski2015kmos3d, 2019ApJ...886..124W}. In agreement, observations with KMOS of 22 star-forming galaxies at z $\sim$ 1.5 with stellar masses 9.5 $\leq $ log(M/M$_{\odot}$) $\leq$ 11.5, found V/$\sigma$ fraction of $\sim $ 1 \citep{2023MNRAS.524.2814P}. In this study, they also derive the outer rotation curves' shape and the dark matter content for these galaxies. They revealed that the dynamics of the rotation-dominated systems is predominantly influenced by dark matter from effective radius scales, in great agreement with cosmological models, with galaxies exhibiting flat or rising rotation curves. In contrast, the dynamics of the dispersion-dominated systems often displayed declining outer rotation curves. 

Several studies \citep{2017Natur.543..397G,2017ApJ...840...92L,2020ApJ...902...98G, price2021rotation,shachar2023rc100} used rotation curves to derive the fraction of dark matter in massive (10$^{10}$ - 10$^{11}$ M$_{\odot}$) star-forming galaxies at z $\sim$ 0.65 - 2.45. The galaxies are rotational-dominated systems with a ratio of V/$\sigma$ varying between 4 and 9. Their observations show a declining rotation curve at larger radii, leading to the conclusion that there is a decrease of the abundance of dark matter in galactic halos at this epoch of the Universe. Extending this finding, \citet{2021sf2a.conf..379B} used data from the Multi-Unit Spectroscopic Explorer (MUSE) and performed a disk-halo decomposition to low-mass (10$^{8.5}$ - 10$^{10.5}$ M$_{\odot}$) star-forming galaxies at z$\sim$ 1. They found that the fraction of dark matter within the half-light radius is $\sim$ 60\% to 95\%, higher than what was reported in \citet{2020ApJ...902...98G}. Although these studies correct for one of the most critical limitations of high-$z$ observations at low spatial resolution - beam smearing, which smooths out velocity gradients and artificially inflate central velocity dispersions \citep{mcgaugh2001high,swaters2009rotation,burkert2016angular} - other observational effects, such as cosmological surface brightness dimming, remain important challenges. Together, these effects can bias the cosmic noon data, making falling rotation curves appear well behaved, as in cold disks, when they are not.

To mitigate the effects of cosmological brightness dimming, \citet{2017ApJ...840...92L} co-added signals from a larger sample of galaxies. They found an averaged declining rotation curve extending out to $\sim$ 4 effective radii. Nonetheless, the effect of interactions and random motion at galaxies outskirts may strongly affect the total signal and final shape of the rotation curve. For example, \citet{2019MNRAS.485..934T} demonstrated that the final shape of the stacked rotation curve is highly sensitive to the normalization method used. When normalizing in size by the galaxy stellar disc-scale length, they recovered the flat galaxy rotation curve extending to 6 disc-scale radius ($R_d$), consistent with predictions from $\Lambda$CDM theory, indicating a dark matter fraction at 6$R_d$ of nearly 60\%. 

Observations of cold molecular gas with ALMA have revealed gas disks with low turbulence and V/$\sigma \approx 2-7$ at cosmic noon \citep{neeleman2020cold,2022A&A...664A..63X, 2023A&A...672A.106L}. In a reanalysis by \citet{2023A&A...672A.106L} of a massive galaxy from \citet{2017Natur.543..397G}, CO line observations showed flat rotation curves and dark matter dominance out to $\sim$ 8kpc, contradicting the \citet{2017Natur.543..397G} result, which pointed to declining rotation curves in the outer regions of the galaxy. Using weak gravitational lensing, \citet{2020Natur.584..201Rizzo} and \citet{2024ApJ...969L...3M} derived circular velocity curves for isolated galaxies at z $\sim$ 6-7 and z $\sim$ 0.2, respectively, finding them to remain flat over hundreds of kiloparsecs, possibly up to 1 Mpc.

While there is now broad agreement on the existence of rotating disks at cosmic noon, discrepancies remain between optical, infrared, sub-millimeter and weak lensing studies. In order to bridge the gap between simulations and observations, \citet{ubler2021kinematics} created a realistic sample of massive galaxies with high star formation rates at redshift $\mathrm{z}=2$, using Illustris TNG50 simulations. Their results indicated that the simulated galaxies were more asymmetric and significantly differed from galaxies observed in the distant universe, highlighting the need to improve the numerical resolution available in current cosmological simulations, as well as more extensive and deeper observations. Studying even more distant galaxies with increased sensitivity and statistical accuracy, while constraining the cosmological bias on these observations, will require next-generation IFU instruments such as MOSAIC on the Extremely Large Telescope \citep{evans2015science}.

In this work, we investigate how the observational bias generated by cosmological surface brightness dimming could affect the cosmic noon data, and we examine how mergers remnants and interacting galaxies that do not have a regular rotation curve would be seen at large redshift.  To achieve this, we use a sample of 19 Lyman Break Analogues (LBAs) \citep[e.g][]{Heckman_2005,hoopes2007diverse} observed in the local universe, comparing them with artificially-redshifted observations, with galaxies shifted to z$\sim$ 2.2. This paper is structured as follows: In Section 2, we present the data and the method used to extract the rotation curves and assess asymmetries. Section 3 outlines a complementary approach we use to derive kinematic asymmetries and presents the main results. In Section 4, we discuss our results, and in Section 5, we summarize our main conclusions.

Throughout this paper we adopt the standard cosmological parameters: $H_0$ = 70 Km s$^{-1}$ Mpc$^{-1}$, $\Omega_m$ = 0.3 and $\Omega_{\Lambda}$ = 0.7.

\section{OBSERVATIONS AND METHODS}
\subsection{Lyman Break Analogs}

Lyman Break Analogues (LBAs) are a class of low-redshift galaxies (z $\sim$ 0.1 - 0.2) that share characteristics with Lyman Break Galaxies (LBGs) at cosmic noon \citep{Heckman_2005}. Defined as a subsample of ultraviolet-luminous galaxies (UVLGs), LBAs possess far-ultraviolet luminosities greater than $2 \times 10^{10} L_{\odot}$ \citep{2007ApJS..173..457B}, which corresponds to the typical rest-frame UV luminosities of LBGs at redshift $\approx 1.5 -3$ \citep{2005ApJ...619..697A,2005ApJ...619L..43A}. This luminosity places them among the most active star-forming galaxies, with a star formation rate up to 100 M$_\odot$yr$^{-1}$ \citep{hoopes2007diverse}. Typically, these galaxies are compact in size, resulting in high surface brightness (I$_{FUV} > 10^9$ L$_\odot$ Kpc$^{-2}$) and display a range of metallicities and dust content, often resembling the properties of cosmic noon LBGs \citep{2009ApJ...699.1307B, 2011ApJ...726L...7O}. Their morphologies can be irregular or disturbed, indicating ongoing interactions or mergers \citep{overzier2010morphologies}. 

These objects present a unique opportunity to evaluate the impact of observational biases at cosmic noon due to their remarkable similarities to typical star forming galaxies at those epochs \citep{basu2009osiris,2010ApJ...724.1373G,overzier2010morphologies}. They show clear signs of irregular morphologies, compact star forming regions and low extinction that are more akin to main-sequence star forming galaxies at $z \sim 2$ than to other galaxies at low redshift, making them the ideal laboratory to investigate in detail physical processes related to galaxy formation that might have occurred frequently 10 billion years ago.  At the same time, \citet{basu2009osiris,2010ApJ...724.1373G,overzier2010morphologies} have shown that many merger signatures, such as tidal tails, asymmetric morphologies, and the presence of companions, become undetectable at $z \sim 2$ due to cosmological surface brightness dimming and reduced physical resolution. This makes such systems an ideal sample for investigating the impact of observational effects on the analysis of rotation curves across different redshifts.

\subsection{Observational data}
The data used in this work were extracted from \cite{2010ApJ...724.1373G}. A total of 21 Lyman Break Analog galaxies (LBAs) were observed from an original sample of 30 LBAs identified by \cite{overzier2010morphologies}. These galaxies were observed using the Keck II telescope at the time equipped with the OSIRIS instrument \citep{2006SPIE.6269E..1AL}. OSIRIS is an integral field spectrograph and utilizes an infrared transmissive lenslet array. The spectrograph has plate scales of 0.020, 0.035, 0.050 and 0.100 arcsec per lenslet and spectral resolution of $\sim$ 3800 (which corresponds to $\sim$ 40 km/s). The instrument is assisted by adaptive optics and operates in two modes: broadband (z, J, H or K) and narrowband. 

A description of the observational data, including galaxy names, redshift (z), AO FWHM (mas), spaxel size (mas), UV half-light radius extracted from HST (kpc), stellar mass (M$_{\odot}$) and star formation rate (M$_\odot$yr$^{-1}$), are listed in Table \ref{tab:Tableobs}. The emission line target across all galaxies was Pa-$\alpha$ (rest wavelength $\lambda$ = 1875.1 nm), redshifted into the middle of the K band (2055 nm $\lesssim \lambda_{obs} \lesssim$ 2350 nm) depending on the galaxy's redshift. A narrowband mode was primarily used to maximize spatial coverage. Most observations were conducted using a 50 mas spaxel scale, covering a field-of-view (FOV) of $\approx 2" \times 3"$; for larger objects, a 100 mas scale was used, effectively doubling the FOV. For further observational details, see \citet{2010ApJ...724.1373G}.

In general, the galaxies were observed for approximately 45 min, achieving a spatial resolution of less than 200 pc. This resolution enabled the detection of companion galaxies, clumps, diffuse emissions, as well as high signal-to-noise ratio (S/N $>$ 6) Pa-$\alpha$ measurements of velocity dispersion and rotation. In contrast, observations of galaxies at redshift $\approx$ 2 using integral field units such as KMOS and SINFONI typically require several hours (4 - 20hr) to achieve a spatial resolution of only a few kiloparsecs (2 - 5 kpcs) \citep{2017ApJ...840...92L, 2017Natur.543..397G}.

\begin{table*}
\centering
\begin{tabular}{|>{\centering\arraybackslash}p{2cm}|
                >{\centering\arraybackslash}p{2cm}|
                >{\centering\arraybackslash}p{2.1cm}|
                >{\centering\arraybackslash}p{2cm}|
                >{\centering\arraybackslash}p{1.5cm}|
                >{\centering\arraybackslash}p{2cm}|
                >{\centering\arraybackslash}p{3.3cm}|}
\hline
Name & Redshift (z) & AO FWHM (mas) & Spaxel (mas) & R$_1$$^a$ (kpc) & log M$_*$ (M$_{\odot}$) & SFR (H$_{\alpha}$ + 24 $\mu$m) (M$_{\odot}$ yr$^{-1}$) \\
\hline
\centering
005527& 0.167& 90 & 50& 0.36 & 9.7 & 55.4\\
015028& 0.147& 82 & 50& 1.34 & 10.3 & 50.7\\
021348& 0.219& 177 & 100& 0.38 & 10.5 & 35.1\\
032845& 0.142& 103 & 50& 0.86 & 9,8 & 8.7\\
035733& 0.204& 116 & 100& 1.00 & 10.0 & 12.7\\
040208& 0.139& 80 & 50& 0.80 & 9.5 & 2.5\\
080232& 0.267& 115 & 100& 3.01 & 10.7 & 30.4\\
080844& 0.096& 187 & 100& 0.08 & 9.8 & 16.1\\
082001& 0.218& 69 & 50& 2.78 & 9.8 & 40.0\\
083803& 0.143& 105 & 50& 1.02 & 9.5 & 6.2\\
092600& 0.181& 101 & 50& 0.68 & 9.1 & 17.0 \\
093813& 0.107& 77 & 50& 0.65 & 9.4 & 19.8\\
101211& 0.246& 96 & 50& N/A & 9.8 & 6.2\\
113303& 0.241& 76 & 50& 1.36 & 9.1 & 7.7\\
135355& 0.199& 68 & 50& 1.45 & 9.9 & 19.4\\
143417& 0.180& 98 & 50& 0.90 & 10.7 & 20.0\\
210358& 0.137& 65 & 50& 0.44 & 10.9 & 108.3\\
214500& 0.204& 70 & 50& 1.13 & 9.9 & 16.4\\
231812& 0.252& 130 & 100& 2.77 & 10.0 & 63.1\\
\hline
\end{tabular}
\caption{Observational data extracted from \citet{2010ApJ...724.1373G}. AO FWHM is the spatial resolution of the image in mas, R$_1^a$ is the UV half-light radius extracted from HST data, the stellar masses were taken from the SDSS/DR7 MPA-JHU catalog \footnote{\url{http://www.mpa-garching.mpg.de/SDSS/DR7/}} and the star formation rates (SFR) were derived from combined H$\alpha$ and MIPS-24µm data and reported by \citep{Overzier_2009}.}
\label{tab:Tableobs}
\end{table*}

\subsection{Data Reduction}

The data were reduced using the OSIRIS pipeline, which performs sky subtraction and transforms the image into a 3D data cube. To create 2D velocity maps, a Gaussian function was fitted to the Pa-$\alpha$ line in each 1D collapsed spectrum. To reduce noise, the data cube was spatially smoothed by 1.5 - 2 pixels, depending on the data quality for each case. This allowed for the detection of emission lines even in the outer regions of the galaxy.

To minimize artifacts in the velocity map, a signal-to-noise (S/N) cutoff of 6 was applied, representing a detection limit of surface density for star formation of $\sum_{SFR} \sim 0.1 M_{\odot}$ year$^{-1}$ kpc$^{-2}$. The S/N ratio was obtained by dividing the area under the Gaussian fit to the Pa-$\alpha$ emission line in each spaxel by the sum of the noise fluctuations in each wavelength interval. The noise was determined in a sky region without any emission lines. 

\subsection{Methodology}

To investigate the reported absence of dark matter claimed by \citet{2017ApJ...840...92L} and \citet{2017Natur.543..397G, 2020ApJ...902...98G}, and to enable a direct comparison between the kinematics of low-redshift galaxies and those at cosmic noon, we simulated cosmic-noon observations by artificially redshifting local galaxies. This approach allows us to isolate and assess the observational biases that may affect cosmic-noon kinematic analyses. Specifically, we used artificially redshifted observations from \citet{2010ApJ...724.1373G}. Altogether, two artificial observations were reproduced, one mimicking the OSIRIS instrument assisted by AO (hereafter, mocked OSIRIS) and another one mimicking the SINFONI instrument without AO (hereafter, mocked SINFONI). The observed galaxies were redshifted to z= 2.2, where this exact value was chosen to avoid major OH emission lines. At this redshift, the galaxy would be observed using the $H_\alpha$ emission line instead of Pa-$\alpha$. Hence, the Pa-$\alpha$ flux maps from actual observations were adjusted to the total $H_\alpha$ flux based on the SDSS catalog \citep{2005AJ....129.1755A}. 

The authors in \citet{2010ApJ...724.1373G} predicted how the Lyman Break Analogs (LBAs) would appear in an observation at a more distant redshift by replicating the IFU prescription described in \citet{law2006predictions}. In this method, the first steps are to rebin the flux maps according to the new angular diameter distance of each galaxy at the new redshift and to convert it into pixel maps, where the pixel dimensions correlate with the desired spaxel scale. The pixel values are adjusted by dividing the total flux of the source by the average photon energy, resulting in a grid of line fluxes for each spaxel. The flux in each spaxel is then approximated by the flux of a point source before incidence on the atmosphere. After crossing both the atmosphere and the adaptive optics (AO) system, the light is split between a diffraction-limited core and a seeing-limited halo. These effects are taken into account using two dimensional Gaussian profiles, with $\sigma_{Halo} = \theta_{seeing}/2.4$ and $\sigma_{core} = (1.22 \lambda/D)/2.95$, where D is the diameter of the telescope's primary mirror and $\lambda$ is the wavelength. The signal from the core and the halo are integrated over each spaxel, generating a flux distribution that incorporates the improvement and losses of the adaptive optics. The surface brightness in each pixel was taken into account according to the relative amount of cosmological dimming of a galaxy at z = 2.2, with respect to the original redshift (i.e., intensity scales as $(1+z)^{-4}$). The background emission was estimated following the Gemini Observatory model for the Mauna Kea NIR sky brightness spectrum\footnote{\url{http://www.gemini.edu/sciops/ObsProcess/
obsConstraints/atm-models/nearIR _ skybg _16 _15.dat.}}, incorporating contributions from zodiacal light (modeled as a 5800 K blackbody), atmospheric emission (modeled as a 250 K blackbody), and radiation from atmospheric OH molecules. Additionally, thermal blackbody emission from warm reflective surfaces along the optical path was included. Noise was introduced in the simulation by propagating both background and source count rates through the optical system. For the mocked OSIRIS setup, the estimated adaptive optics (AO) Strehl ratios were also taken into account. Mock integration times are between 2 and 3 hours to emulate typical cosmic noon observations such as those found in \citet{law2006predictions} and \citet{Forster}, and the pixel scale and spectral resolution match the ones in those works.

In the mocked SINFONI observations, the instrument was considered to exhibit greater sensitivity, as there is no loss due to the AO Strehl system. Sensitivities in terms of surface brightness is the same in seeing-limited observations at cosmic noon and AO-assisted observations at $z\sim 0.2$, down to 0.1 M$_\odot$ yr$^{-1}$ kpc$^{-2 }$, due to a combination of the increased luminosities of H$_\alpha$ compared with Pa-$\alpha$, no loss to strehl and the opposing effect of decreased signal due to cosmic dimming. Therefore, the observation was simulated by solely rebinning the datacubes to the SINFONI pixel scale (0.125") while simultaneously degrading the spatial resolution using a Gaussian kernel of 0.5". The total angular size of the galaxy was adjusted in accordance with the angular diameter distance at $\hbox{z} \sim 2.2$ and the original redshift of each galaxy. It is expected that the decrease in surface brightness is offset by the increased signal resulting for the instrument’s higher sensitivity; therefore, no extra corrections for cosmological dimming were applied.

\section{RESULTS}

We extract the kinematics of our galaxies and use two approaches to identify signs of interactions and perturbations in the velocity field: The first involves extracting velocities along the kinematic position angle (PA) of each galaxy, constructing rotation curves and comparing both sides of it by fitting a polynomial function using the same method described in \citet{ubler2021kinematics}, while the second utilizes Kinemetry, a software for analyzing kinematics of 2D IFU maps, extracting rotation curve and detecting deviations from disk-like motions following the method described in \citet{2008ApJ...682..231S}. The methods used in this work operate in two dimensions, relying on velocity fields rather than full data cubes. Consequently, they do not correct for beam smearing effects as accurately as three-dimensional forward modeling techniques (e.g., GALPAK-3D \citep{bouche2015galpak3d}, 3D-BAROLO \citep{teodoro20153d}, and BLOBBY-3D \citep{varidel2019sami}). However, our goal in this work is not to recover the intrinsic kinematics in full detail, but rather to replicate the typical methodology employed in many cosmic-noon studies, which rely on two-dimensional kinematic maps.

\subsection{Rotation Curves}
\label{rc}
We extract rotation curves for 15 objects. Galaxy 080232 was not resolved even with adaptive optics; therefore, we exclude it from our analyses. Four of the mocked OSIRIS galaxies (113303,101211,080844,040208) were too small - less than 3 points in each side of the rotation curve - therefore they were also excluded from the analyses. 

\begin{figure}[h!]
    \centering
    \includegraphics[width=1\linewidth]{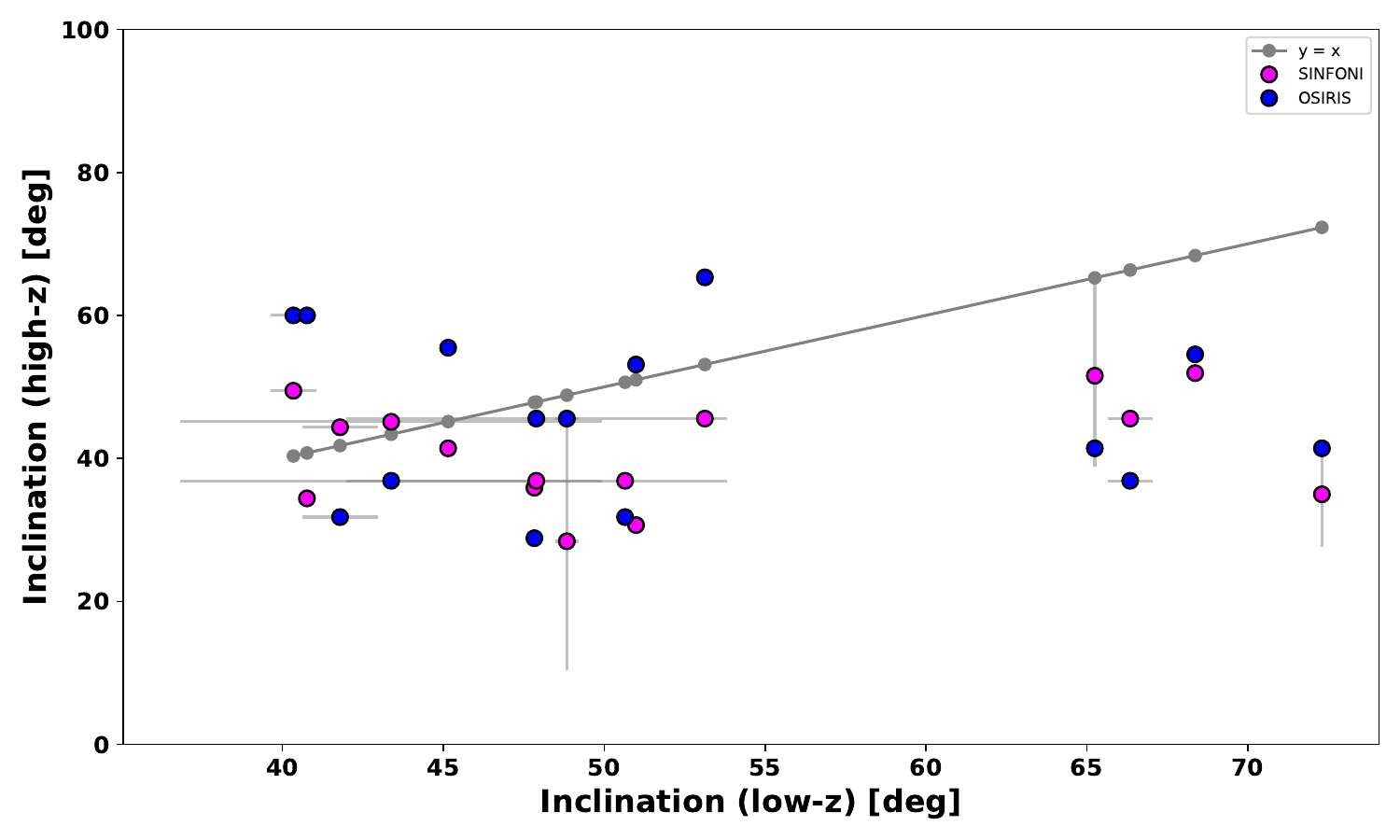}
    \caption{Inclination estimated in low redshift galaxies versus inclination in cosmic noon galaxies. The blue dots refers to the mock observations made with OSIRIS using AO and the pink dots represents the mock observations with SINFONI. The gray line illustrates what we would observe if the inclination between both redshift regimes was the same.}
    \label{inc}
\end{figure}

We estimate the inclination of each galaxy using \textsc{find-galaxy}\footnote{\url{https://pypi.org/project/mgefit/\#mge-find-galaxy}}, assuming that the gas is distributed in a disk. We corrected the observed velocity by deprojecting the line-of-sight component using the derived inclination angle. The deprojected velocity is calculated as:

\begin{equation}
    V_{\text{rot}} = \frac{V_{\text{los}}}{\sin(i)},
\end{equation}

\noindent where $V_{\text{rot}}$ is the intrinsic rotational velocity, $V_{\text{los}}$ is the observed line-of-sight velocity, and $i$ is the inclination angle. In general, we find that the inclinations of high-redshift galaxies appear smaller than those of their low-redshift counterparts, meaning that the simulated galaxies appears more face-on. This is consistent with expectations, considering the blurring effect of seeing-limited observations and the loss of information in galaxy's outskirts (see Fig.~\ref{inc}).

Due to the irregularity of the galaxies, the kinematic center and the rotation axis are not always well defined; therefore, we use \textsc{PaFit}\footnote{\url{https://www-astro.physics.ox.ac.uk/~cappellari/software/\#sec:pafit}} - presented in Appendix C of \citet{krajnovic2006kinemetry} to extract the kinematic center and the position angle (PA) of all galaxies. 
In essence, the software measures the global kinematic position angle of the two-dimensional projection of velocities and determines the position of the apparent angular momentum of a galaxy ($\boldsymbol{L_p}$). The apparent angular momentum can be defined as a projection of the intrinsic angular momentum to the plane of the sky:

\begin{equation}
\centering
    \boldsymbol{L_p} = \int_P \boldsymbol{r_p} \times \Sigma\boldsymbol{v_r} d\boldsymbol{r_p}
\end{equation}

\noindent where $\boldsymbol{r_p}$ is a projection of the vector $\boldsymbol{r}$ onto the plane of the sky, at which the mean velocity of the stars (v) projects to the observed radial velocity vector $\boldsymbol{v_r}$. $\Sigma$ represents the surface density and the integration is over the entire projection plane.

Since the apparent angular momentum ($\boldsymbol{L_p}$) is entirely determined by the observed surface brightness and the two-dimensional velocity map of the galaxy, it simplifies the interpretation of the galaxy’s dynamics. For an axisymmetric galaxy, when the figure rotation is absent, $\boldsymbol{L_p}$ aligns with the projection of the intrinsic angular momentum, reducing the complexity of the three-dimensional velocity space to the projection of the angular momentum \citep{Franx1988}. Thus, by measuring the luminosity-weighted kinematic position angle (PA) from the velocity map, one can determine the position angle of the apparent angular momentum. 

To ensure a more robust extraction of the kinematics for each object, we utilize \textsc{PaFit} exclusively within the central velocity structure associated with the main galaxy, excluding companions and avoiding artifacts. Correspondingly, the maximum and minimum velocity values are defined as the averages of the upper and lower five percent of values, respectively, enhancing the accuracy of the kinematic measurements. The five percent threshold was adopted to ensure consistency with the methodology used in \citet{2010ApJ...724.1373G}, where the authors verified that this criterion effectively excluded outliers without removing spaxels that followed the overall kinematic behavior of the galaxy. To estimate the kinematic center, we use a grid-search approach around an initial guess based on the galaxy centroid extracted from the flux map. We systematically test possible center positions within ±2 pixels in both the x and y directions. For each trial center and position angle (PA), we reconstruct a model velocity field, assuming it is symmetric, and compare it to the observed velocity field. The best-fit center is the one that minimizes both the chi-square value and the systemic velocity, providing a refined estimate of the kinematic center.

\begin{figure*}
    \centering
    \includegraphics[width=1\textwidth]{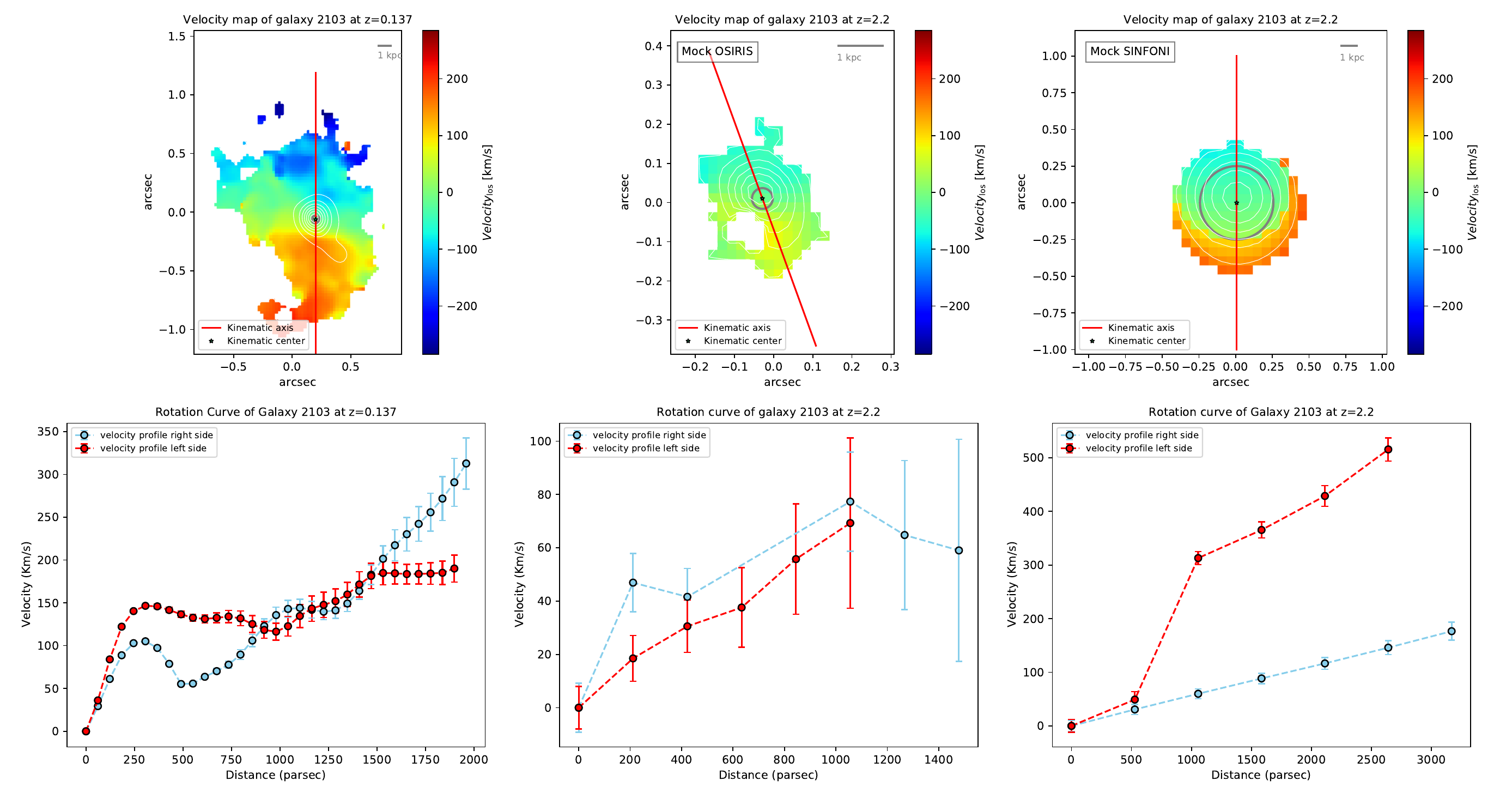}
    \caption{Top: From left to right velocity map of galaxy 210358 extracted from (original) Pa-$\alpha$ observations using OSIRIS instrument assisted by AO, followed by velocity map extracted for mocked OSIRIS H$\alpha$ observations at z$\sim$2.2, assisted by AO and lastly velocity map extracted for mocked SINFONI H$\alpha$ observations at z$\sim$2.2, without AO. The axes display the angular scale in arcseconds, maintaining a consistent orientation across all panels, with north pointing up and east to the left. Each panel shows the FWHM of a point source, represented by a grey circle centered on the galaxy, as an indicator of spatial resolution, along with a bar indicating the physical scale corresponding to 1 kpc at the galaxy’s redshift. The red line represents the kinematic axis used to extract the rotation curves. White contours represent the signal-to-noise distribution of the emission line map. Bottom: Rotation curve corrected for inclination extracted along the kinematic axis for original Pa-$\alpha$ observations(left), followed by mocked OSIRIS observation at z $\sim$ 2.2 (middle) and lastly mocked SINFONI observation at z$\sim$ 2.2 (right). 
    }
    \label{2103map}
\end{figure*}

In Figure~\ref{2103map}, we show, as an example, the velocity maps of galaxy 210358, each overlaid with S/N contours and the major kinematic axis (top panels), together with the corresponding rotation curves extracted along the major kinematic axis (bottom panels). To enable a direct comparison between both sides of the galaxy, the sign of the approaching side was flipped with respect to the galaxy center, and both sides are plotted together.

In general, we observe that the original OSIRIS velocity maps are characterized by complex structures that differ from those of idealized rotating disks. This complexity is reflected in rotation curves with multiple peaks and troughs, which may be indicative of non-circular motions introduced by galaxy interactions or other dynamical disturbances. In contrast, the redshifted mock observations tend to exhibit much smoother rotation curves. This effect is more prominent in the mocked SINFONI observations and occurs because the inner regions of the galaxies become blended due to beam smearing (see Fig.~\ref{2103map}, top-right panel, for an example).

In addition to the smoothing of the inner regions, we also observe a loss of information regarding the high-velocity values in the galaxy outskirts, primarily due to surface brightness dimming, as shown in the velocity map and rotation curve of galaxy 210358 in Fig. \ref{2103map}. This phenomenon is more pronounced in the mock observations with OSIRIS using adaptive optics, due to the fact that the Strehl system exacerbates the loss of sensitivity in the outer regions.

The combined effects of resolution loss at the central regions and the loss of information in the galaxy outskirts can result in a dramatic shift in the estimated kinematic center and position angle of between observed and mocked observations (e.g. Fig. \ref{2103map}). Even when the galaxy does not have a well defined velocity gradient neither kinematic position angle (PAkin), the cosmic noon mocked observations can artificially get a better defined gradient and PAkin, potentially misleading observers to interpret it as a much simpler rotating system.

\subsection{Asymmetries}

It is well known that mergers and interactions can disturb the gravitational potential, especially in the outer parts of the galaxies, leaving imprints of asymmetries in their kinematics, see for example the asymmetry between both sides (red and blue) of the rotation curve of galaxy 210358 in Fig \ref{2103map} (bottom left). We quantify the asymmetry of the rotation curves in the real galaxies and compare it with mocked observations of the same objects in distant universe. To account for the asymmetries, we replicate the method of \citet{ubler2021kinematics}. While this method provides a simple, quantifiable metric, it is important to note that fitting quadratic functions to rotation curves do not accurately represent the true kinematic structure of galaxies, particularly in dynamically complex systems such as LBAs. Nonetheless, we adopt this method to enable direct comparison with the results of \citet{ubler2021kinematics}, which facilitates a consistent evaluation of how observational limitations affect both samples. The method involves fitting a quadratic function to one side of the rotation curve, computing the reduced chi-squared $\chi^2_{\text{red}}$ between this fit and the opposite side of the curve, and then repeating the same process in reverse. 

Considering a model $f$ with parameters $\vec{\Theta}$ and $N$ data points $y_n$, the $\chi^2_{red}$ is given by:
\begin{center}
    \begin{equation}
        \chi^2_{red} = \frac{1}{K}\sum_{n=1}^N \left( \frac{y_n -  f(\vec{x_n};\vec{\Theta})}{\sigma_n}\right)^2,
    \end{equation}
\end{center}

\noindent where $\sigma_n$ are the Gaussian errors at each position $x_n$ and the $K$ are the degrees of freedom. 
To estimate the errors, we used Monte Carlo approach, perturbing the velocity measurements by their respective uncertainties across 100 iterations. For each perturbation, a polynomial model was fitted to the perturbed data and the corresponding reduced $\chi^2$ was calculated, yielding a distribution that reflects the fluctuations due to measurement noise. The final error was estimated by calculating the standard deviation of this distribution.

\begin{figure}
\includegraphics[width=1\linewidth]{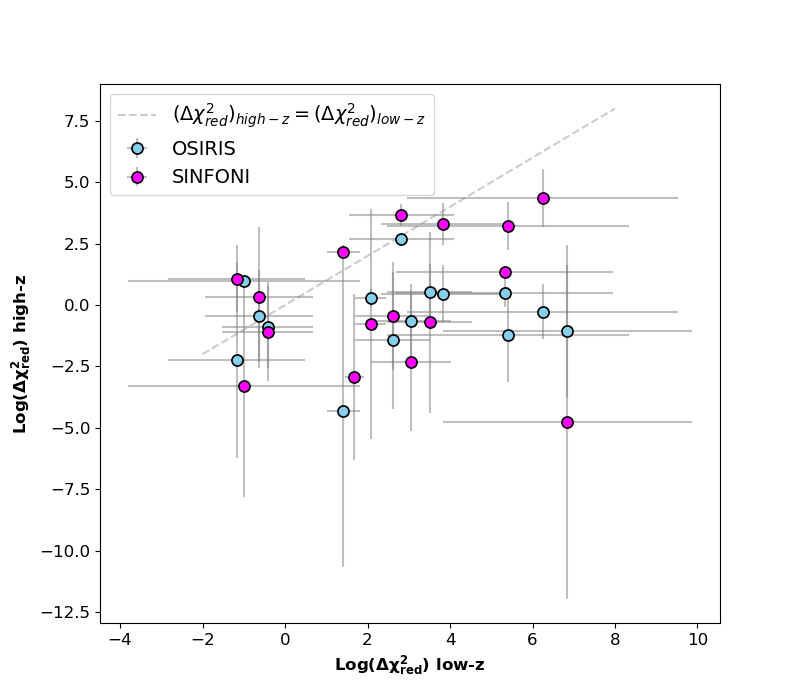}
    \caption{Measures of asymmetry in low redshift galaxies versus asymmetry in cosmic noon galaxies. The blue dots refers to the mock observations made with OSIRIS using AO and the magenta dots represents the mock observations with SINFONI. The gray line illustrates what we would observe if the asymmetry between both redshift regimes was the same.}
    \label{logchi}
\end{figure}

We first fit quadratic curves to each side of the rotation curve, then, we find the $\chi_{\text{red}}^2$ from eq. 2 between a model of one side of the rotation curve and the measured other side of the rotation curve. The same process is repeated using the other rotation curve model-data pair. Finally, we obtain $\Delta \chi_{\text{red}}^2$ by computing the average between the goodness-of-fit for both sides of the rotation curve.
In Figure \ref{logchi}, we present the resultant $\Delta \chi_{\text{red}}^2$ contrasting the real and mocked observations for all galaxies. In the figure, we represent the mocked observation with OSIRIS instrument in blue and magenta for the mocked observations with SINFONI instrument. 
We observe that in the majority of the cases, the values of $\Delta \chi_{\text{red}}^2$ are smaller in the cosmic noon samples (both OSIRIS and SINFONI) than for the low-z sample. This means that the cosmic noon mock galaxies appear to be more symmetric than their real progenitors.

\subsection{Kinemetry}

We also use Kinemetry as an additional method to estimate the asymmetries of rotation curves. Kinemetry \footnote{\url{https://www.aip.de/en/members/davor-krajnovic/kinemetry/}} is a software for analyzing 2D maps of velocity distributions obtained along the line of sight with integral field spectroscopy. The software operates by performing a harmonic expansion of observable quantities, such as velocity, along the ellipses that best fit the 2D maps. The method was developed by \cite{krajnovic2006kinemetry} and its main objective is to quantify the components of the moment maps and to detect and describe morphological and kinematic structures.

Considering a velocity map K(a, $\psi$), one can divide the velocity profile into elliptical rings and describe them with a finite number of harmonic terms in a Fourier expansion:

\begin{equation}
    K(a, \psi) = A_0(a) + \sum^N_{n=1} k_n (a)cos[n(\psi - \phi_n(a))],
\end{equation}

where $\psi$ is the azimuthal angle measured from the major axis in the galaxy plane, $a$ is the major axis of the ellipse and the amplitude and phase coefficients (k$_n, \phi_n$) can be calculated from the sine coefficients (A$_n$) and the cosine coefficients (B$_n$) as:
\begin{equation}
    k_n = \sqrt{A_n^2 + B_n^2} \quad\hbox{and}\quad \phi_n = \hbox{arctan} \left( \frac{A_n}{B_n} \right),
\end{equation}

The best-fitting ellipse is obtained by minimizing the terms of the expansion. To determine the radius from which we will extract velocity values, we impose a conservative condition that requires 95\% of the points along an ellipse to be present in the velocity map. 

While we acknowledge that 2D methods are limited in capturing the full kinematic complexity, especially in low-resolution data where beam smearing is significant, our focus here is not on recovering the exact intrinsic kinematics. Instead, we aim to evaluate how observational effects influence the irregularities presents in the rotation curves of the LBAs in low redshift, using analysis methods commonly adopted in cosmic-noon studies. In this context, we run kinemetry choosing a number of 10 terms for the expansion to measure the asymmetric components. We estimated an initial value for the galaxy's center and position angle by using PaFit as described in Section \ref{rc} and used that value as input for the algorithm. For a velocity field in an ideal, rotating disk, the only non-zero kinemetry coefficient ought to be $B_1$, as this term is the rotation curve for each galaxy. 

\begin{figure*}
    \centering
    \includegraphics[width=1\linewidth]{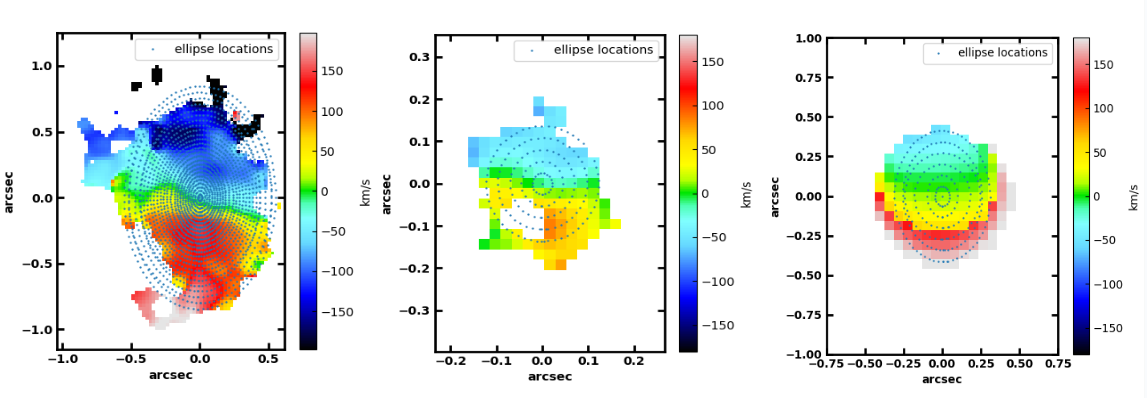}
    \caption{Velocity maps of the galaxy 210358 with blue dots representing the ellipses used in Kinemetry to extract the rotation curves.Left: Real data, Middle: Mock observation with OSIRIS at z$\sim$ 2.2. Right: Mock observation with SINFONI at z $\sim$ 2.2.}
    \label{velkinmaps}
\end{figure*}

Information about additional asymmetries is then contained in the higher-order terms of the expansion. Our second method to quantify the asymmetry in the rotation curves is based on \citet{2008ApJ...682..231S}. In this method, the asymmetry is calculated by performing an average of the higher-order terms of the Fourier expansion $K_{avg,v} = K_2 + K_3 + K_4 + K_5$, subsequently normalizing the average value by the dominant harmonic coefficient $B_1$ to assess the relative level of deviation for each object. Therefore:
\begin{equation} 
v_{asym} = \left\langle \frac{K_{avg,v}}{B_1} \right\rangle_r
\end{equation}

This approach enables us to identify and quantify components in the velocity field that deviates from an ideal rotating disk. Given that mergers and interactions can produce highly disturbed velocity fields, both even and odd terms of the expansion are included in this analysis.  Before taking the final average of $v_{asym}$ across all radii, we construct galaxy's asymmetry curve along its radius. Subsequently, we use error propagation to determine the uncertainties associated with the asymmetry along each curve. 

\begin{figure*}[h!]
    \centering
        \includegraphics[width=0.75\linewidth]{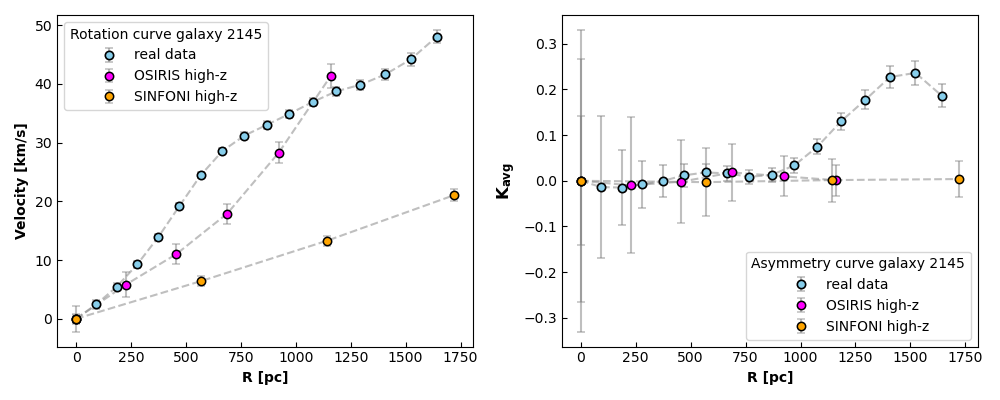}
    \caption{Left: Rotation curve of galaxy 214500 extracted using Kinemetry. Right: Asymmetry curve extracted by summing the highest terms in the Fourier expansion. In both curves, the blue dots represents the real observations at z $\approx 0.2$, the pink dots indicates the mocked OSIRIS observation at z $\approx 2.2$ and the orange dots represents the mocked observation with SINFONI.}
    \label{kincrv}
\end{figure*}

\begin{figure*}[h!]
    \centering
        \includegraphics[width=0.75\linewidth]{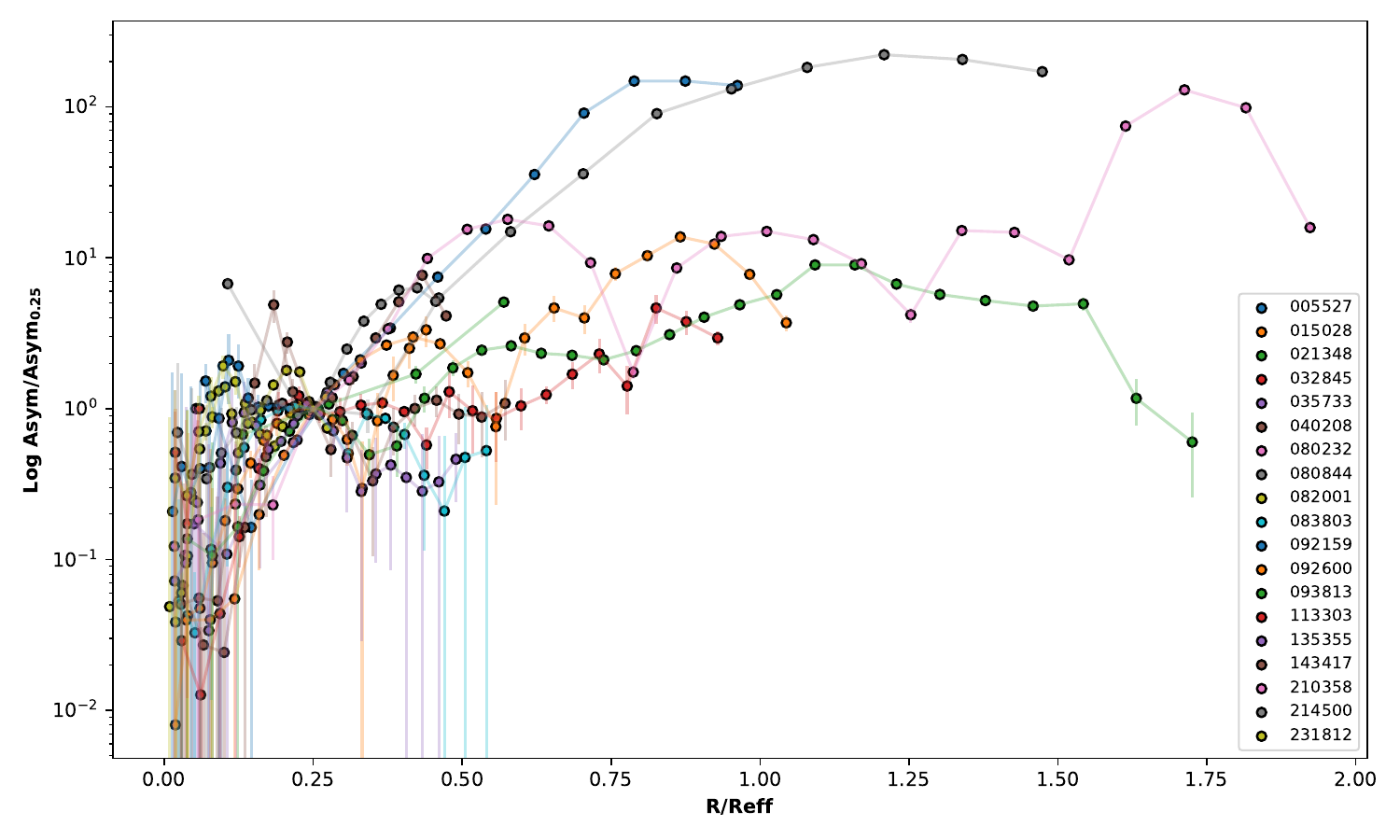}
    \caption{Asymmetry of all LBAs normalized by its value at R/Reff = 0.25, plotted in function of the radius, normalized by the UV half-light radii obtained from HST data. Each color represent an LBA, with its name indicated in the legend. }
    \label{asym}
\end{figure*}

Using Kinemetry, we were able to extract the rotation curve, asymmetry and dispersion over the entire LBAs radius. In Fig. \ref{kincrv}, we show a combined plot with real and mocked data for galaxy 214500, highlighting the difference between the three curves. In this analysis, we measured that the asymmetry in the original observations increases towards the galaxy's larger radius and becomes significant in the outskirts with a Spearman mean $\rho$ correlation of 84\% and Pearson 77\%.
This effect can also be seen in Fig. \ref{asym} where we show the asymmetry for all galaxies in the sample normalized by its value at R/Reff = 0.25. 
The fact that the asymmetry grows with galaxy radius can be interpreted as a consequence that the impact of the interactions between galaxies are stronger at larger radii. 

\begin{figure}[h!]
    \centering
    \includegraphics[width=1.1\linewidth]{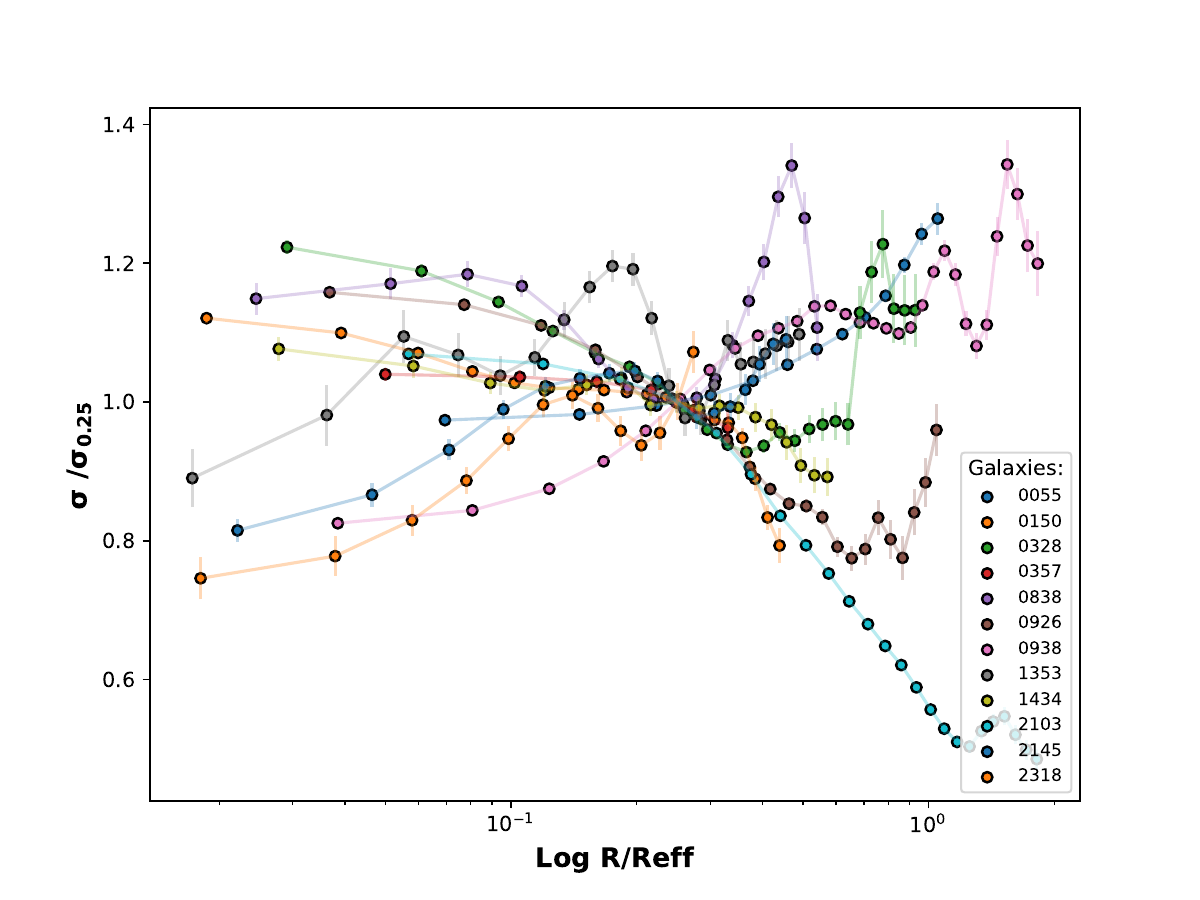}
    \caption{ Velocity dispersion curves for all LBAs, normalized by their values at \( R/R_{\text{eff}} = 0.25 \), are shown as a function of radius normalized by the UV half-light radii derived from HST observations. Each curve corresponds to a different LBA, with colors indicating individual galaxies as labeled in the legend.}
    \label{disp}
\end{figure}

The impact of processes occurring in the outskirts of galaxies is also evident in the dispersion curves of the local LBAs (see Fig. \ref{disp}). In this plot, we present the velocity dispersion profiles of the galaxies at z $\sim$ 0.2, normalized by their values at R/R$_{eff}$=0.25. We observe that in some cases, the dispersion is constant or increases in the outer regions, which may be attributed to low signal-to-noise data. Additionally, external perturbations, possibly caused by nearby companion galaxies, may contribute to the dynamical heating in these areas.

\begin{figure}
    \centering
        \includegraphics[width=0.95\linewidth]{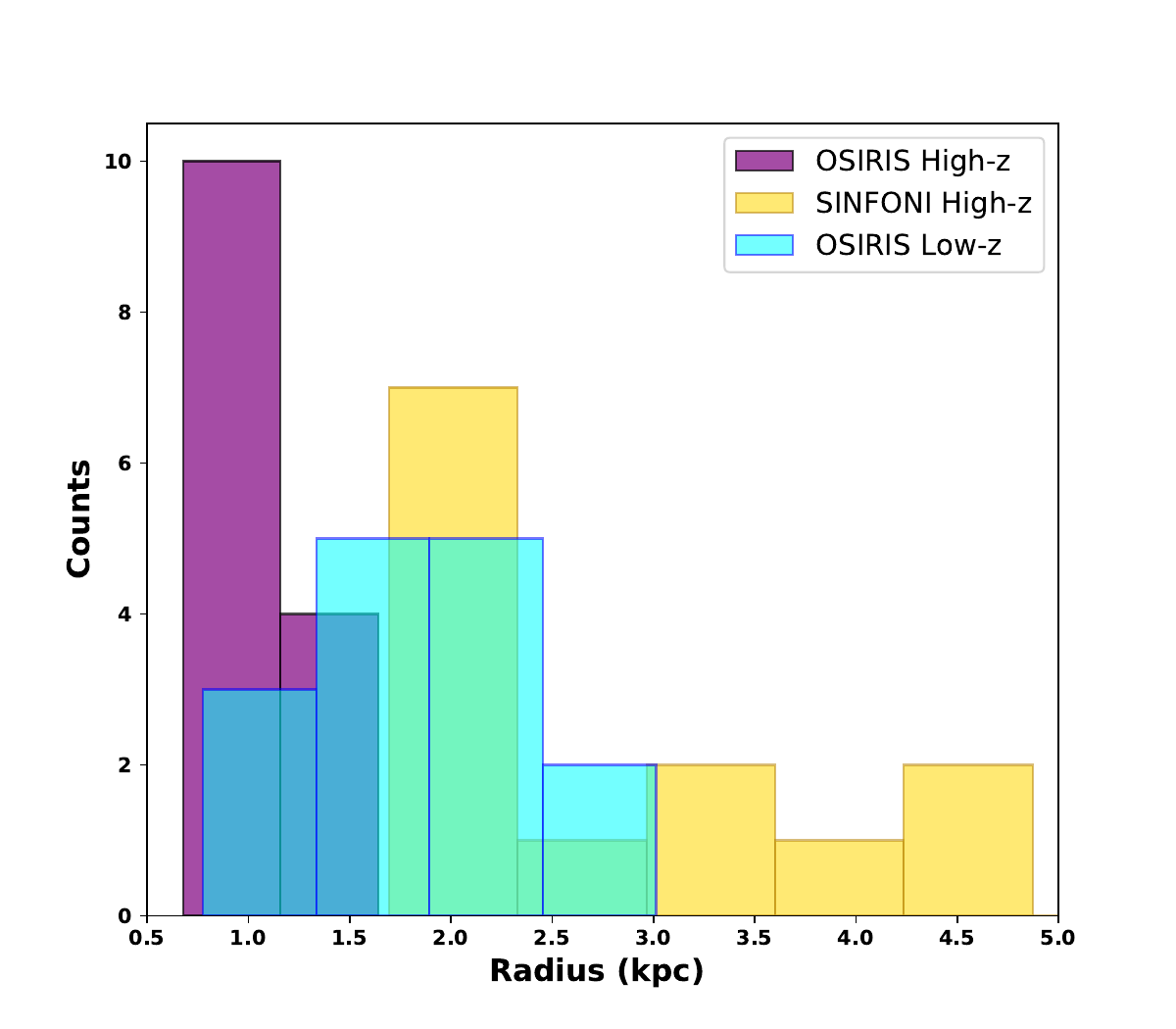}
    \caption{Maximum radius of the rotation curve for each galaxy achieved in the real (light blue) and mocked observations (OSIRIS - purple, SINFONI - yellow). }
    \label{sizes}
\end{figure}

\begin{figure}
    \centering
    \includegraphics[width=0.99\linewidth]{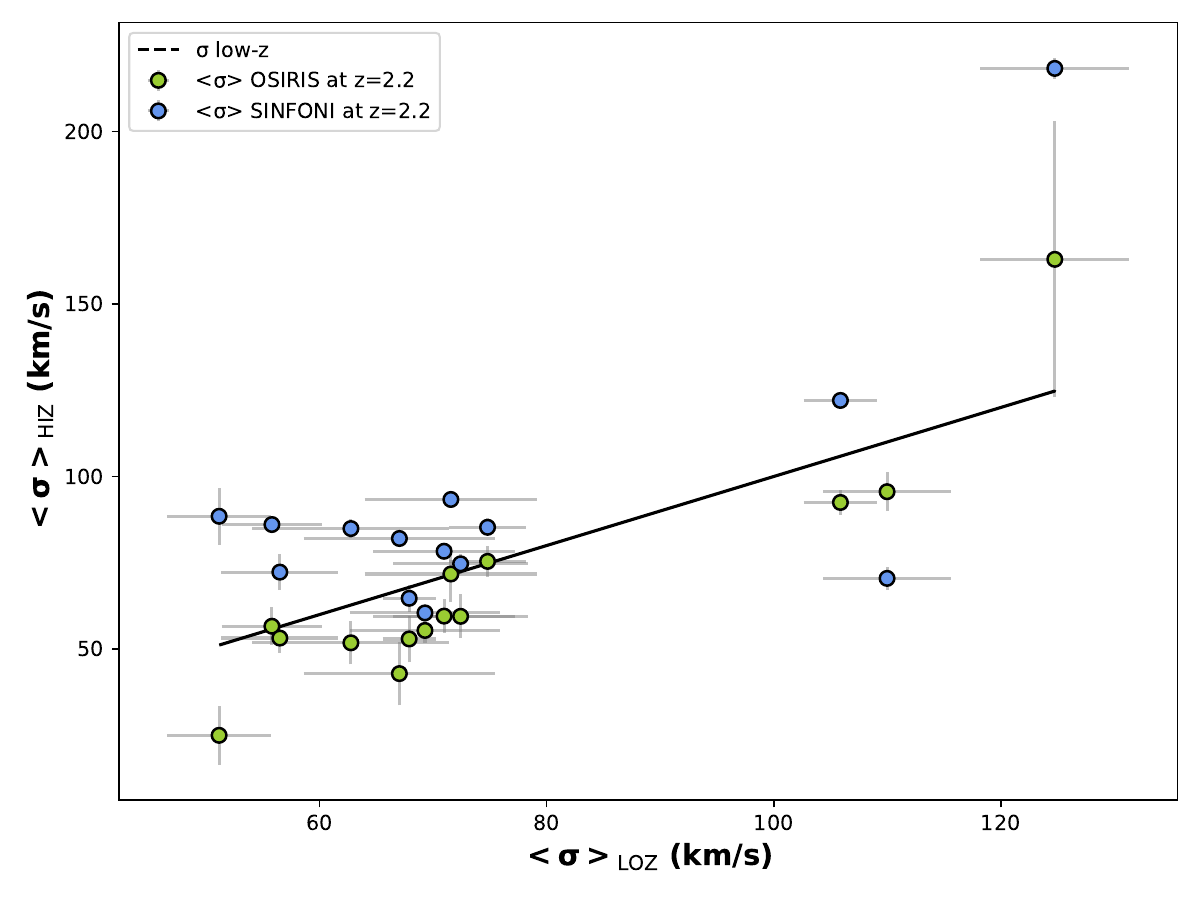}
  \caption{Measurements of mean velocity dispersion in low-redshift galaxies compared to those at cosmic noon, extracted using Kinemetry. Blue dots represent the mock observations with OSIRIS, while green dots correspond to mock observations with SINFONI.}

    \label{dispmean}
\end{figure}

At cosmic noon, we are unable to observe the outer regions of these galaxies due to cosmological surface brightness dimming. Figure \ref{sizes} illustrates the difference in length of rotation curve between real and mock observations, as can also be seen in the example shown in Figure \ref{kincrv}. The simulated curves are, on average, 1.5 times smaller than the real ones for the OSIRIS instrument assisted with AO. The SINFONI instrument is more sensitive than OSIRIS with AO and can capture the external parts, but its loss of resolution tends to blur the asymmetric features. The effect of beam smearing is illustrated in Figure ~\ref{dispmean}, where we show the median velocity dispersion values for both local and mock observations. The impact of low resolution is more significant in the mock observations with SINFONI, since they lack adaptive optics; as a result, the mean dispersion of the galaxies appears higher than their low-redshift counterparts. On the other hand, mock observations with OSIRIS show lower dispersion values, as the galaxies are more compact and do not display additional high-dispersion values in their outskirts.

\begin{figure}
    \centering
    \includegraphics[width=1\linewidth]{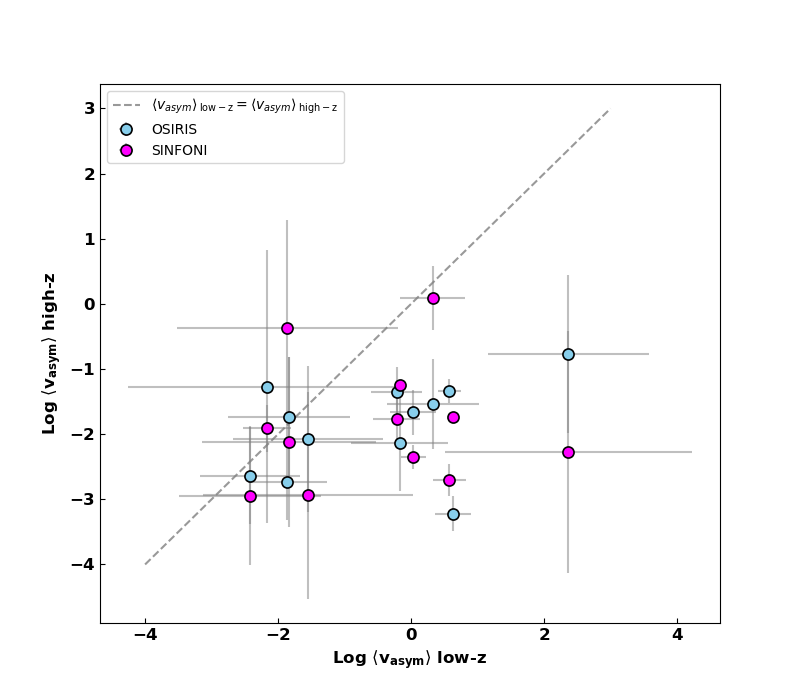}
    \caption{Measures of Asymmetry in low redshift galaxies versus asymmetry in high redshift galaxies using the method described by \citet{2008ApJ...682..231S}. The blue dots represents the artificially-redshifted observations with OSIRIS and the pink dots the artificially-redshifted observations with SINFONI. }
    \label{asy}
\end{figure}

In Fig. \ref{asy}, we show the asymmetry in low-z galaxies versus the asymmetry in cosmic noon derived using the \citet{2008ApJ...682..231S} method. The mocked observations 
are represented in blue dots for OSIRIS instrument and in magenta for SINFONI. Once again we found that the mean value of asymmetry (v$_{asym}$) is smaller at cosmic noon than in the local universe. 

\begin{figure} [h!]
    \centering
    \includegraphics[width=1\linewidth]{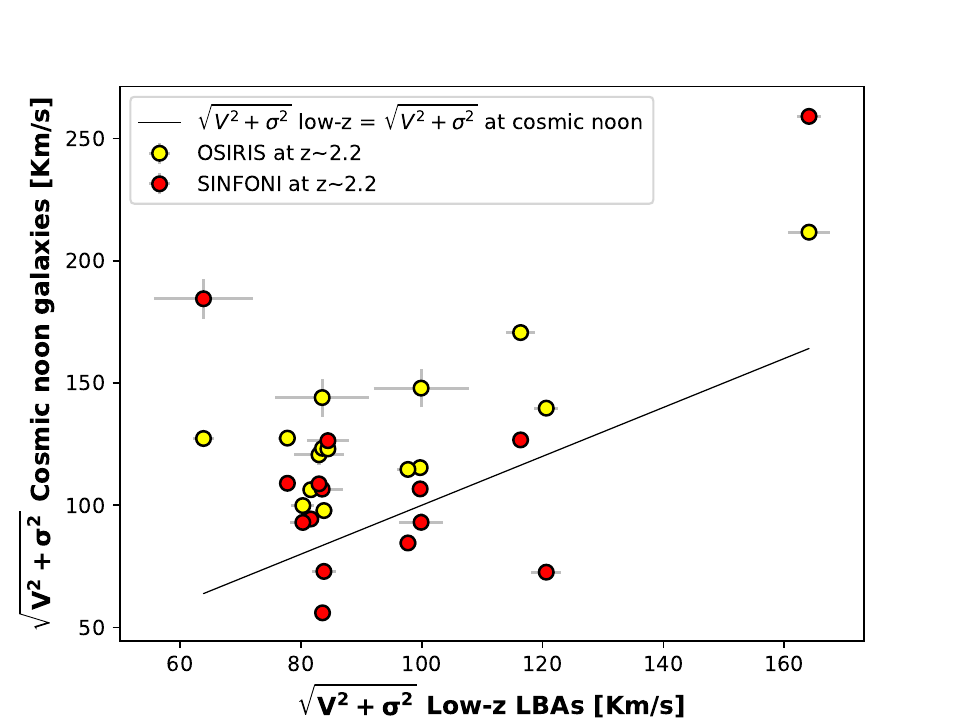}
    \caption{Measures of second order velocity moment corrected for inclination V$_{rms} = \sqrt{V_{rot}^2 + \sigma^2}$ between real and mocked observations - OSIRIS (yellow) and SINFONI (red). The black solid line corresponds to the equally kinematic systems.}
    \label{vsi}
\end{figure}

We assessed the kinematic similarity between the original and artificially redshifted data by computing the second-order velocity moment, \(V_{\mathrm{rms}}\), as a proxy of the gravitational potential energy for each observation. Following \citet{binney2005rotation}, \(V_{\mathrm{rms}}\) corresponds to the quadratic sum of the observed rotational velocity (\(V\)) and velocity dispersion (\(\sigma\)). When properly weighted by mass, this quantity represents the total kinetic energy along the line of sight. Since kinetic and potential energies are connected through the virial theorem, \(V_{\mathrm{rms}}\) is directly related to the gravitational potential energy. In \citep{2017A&A...608A...5G} was shown that while \(V_{\mathrm{rms}}\) is well defined for collision-less systems, i.e. stars, the gas follow (approximately) 1:1 relation with the stars. We therefore  approximate \(V_{\mathrm{rms}}\) as \( \sqrt{V^2 + \sigma^2} \), and measure \(V_{\mathrm{rms}}\) by summing over all spaxels with \(\mathrm{S/N} > 6\) in the kinematic map of each galaxy. We present the result in Fig. \ref{vsi}, where we plot the V$_{rms}$ of mocked OSIRIS (yellow) and mocked SINFONI (red) as a function of the V$_{rms}$ of low-z original data. Figure \ref{vsi} reveals that the V$_{rms}$ of the mocked galaxies does not follow a 1:1 relation with the V$_{rms}$ of the original LBAs observations (black line). This indicates that galaxies observed at cosmic noon appear kinematically different from their true low-\(z\) counterparts and do not trace the same gravitational potential. Since both data sets correspond to the same galaxies, this discrepancy highlights the significant impact of observational effects on the inferred kinematics.

\section{DISCUSSION:}

The dynamics of stars and gas within galaxies are mainly influenced by the gravitational potential, determined by the mass distribution of dark matter and baryons. When comparing the ratio of rotation to dispersion velocities (V/$\sigma$), it was shown that cosmic noon star-forming galaxies exhibit higher gas turbulence in rotating disks compared to their lower redshift counterparts \citep{basu2009osiris,2019ApJ...886..124W,2023MNRAS.525.2789D}. However, the detailed observation required to study dark matter content and the drivers of turbulence in cosmic noon galaxies are limited by the low quality of recent data. \citet{rizzo} emphasized that high angular resolution and high signal-to-noise ratio (SNR) are crucial to accurately assess the dynamical state of galaxies. Although many studies highlight velocity gradients in distant galaxies \citep[e.g][]{10.1093/mnras/stab2226}, it remains unclear whether these gradients indicate regularly rotating discs or poorly resolved, disturbed systems. This uncertainty is mainly compounded by cosmological surface brightness dimming and the loss of resolution in distant objects, which often cause the outer regions of galaxies to vanish, leaving behind a smaller disk-like structure. For instance, \citet{Swaters_2000} showed that H$\alpha$ rotation curves often display different shapes compared to those derived from HI observations, likely because H$\alpha$ traces the gravitational potential less accurately than HI, due to its higher temperature and more limited spatial extent.


Observations of warm ionized gas have revealed massive, star forming galaxies with v/$\sigma \sim 4-9$ at z $\sim$ 0.6 - 2 \citep{2017ApJ...840...92L, 2017Natur.543..397G, 2020ApJ...902...98G, Genzel_2020,price2021rotation,shachar2023rc100}. These studies extracted rotation curve and estimated dark matter content for each galaxy. Considering that the galaxy's rotation curve declined at larger radii, these studies concluded that there is a relative decrease in dark matter abundance at this epoch, based on comparisons with simulations. However, surveys such as $KMOS^{3D}$, which include lower-mass galaxies (down to $\sim 10^9 M \odot$), often report lower v/$\sigma$ values, closer to unity. A recent study using JWST \citep{danhaive2025dawn} also revealed a population of turbulent disks with v/$\sigma \sim 1–2$. Found at significantly higher redshifts (z $\sim$ 3.6–6.5) these galaxies span a broader mass range ($\sim 10^8 - 10^{10} M\odot$),with the most rotation-dominated systems corresponding to the higher-mass end ($\sim 10^9 - 10^{10} M\odot$). Nonetheless, as shown in observations of nearby H$\alpha$ emission galaxies using Sydney–AAO Multi-object IFS (SAMI) survey \citep{10.1093/mnras/sty273} and Mapping nearby Galaxies at Apache Point Observatory (MANGA) survey \citep{Feng_2022}, the dynamics of the outer regions of galaxies are highly sensitive to mergers and interactions. These factors, combined with observational limitations, make it challenging to accurately recover the true gravitational potential at this epoch.

To account for kinematic disturbances from mergers and interactions, we apply two methods. The first method we use to assess asymmetry is described in \citet{ubler2021kinematics}. In this approach, we fit a quadratic function to each side of the rotation curve and compute the reduced chi-squared between one side’s quadratic model and the measured data from the opposite side. This process is repeated for the other side, yielding the $\Delta\chi_{red}^2$ for each galaxy. Applying this method to original observations of LBAS, we found mean values of $\Delta\chi_{red}^2 = 158.05$ at z $\sim$ 0.2 and $\Delta\chi_{red}^2 = 2.81$ for mocked observations made with OSIRIS and $\Delta\chi_{red}^2 = 33.04$ for SINFONI at z $\sim$ 2.2. In comparison, \citet{ubler2021kinematics} extracted $\Delta\chi_{red}^2$ for five mock observations of galaxies extracted from TNG50 and twelve massive observed galaxies from \citet{2017Natur.543..397G, Genzel_2020}. They report mean values of $\Delta\chi_{red}^2 = 46.8$ for mock observations of TNG50 simulated galaxies at z $\sim$ 2 and $\Delta\chi_{red}^2 = 2.1$ for observed galaxies from \citet{2020ApJ...902...98G} at cosmic noon. These results align with our findings, where observed galaxies at cosmic noon exhibit smaller $\chi_{red}^2 $ values, indicating that distant objects appear to be more symmetrical than they really are. 

Second, we quantify asymmetry by calculating deviations from circular motion in velocity maps, using higher-order terms in a Fourier Series  \citep[e.g][]{krajnovic2006kinemetry,2008ApJ...682..231S,2023ApJ...957...48G}. With Kinemetry, we derive rotation and asymmetry curves for each galaxy, finding that interactions have a pronounced impact on the outer regions, where the curves becomes highly asymmetric. In contrast, when comparing this with mocked observations of the same objects placed at cosmic noon, we observe that this asymmetry is less prominent in cosmic noon data. In mocked observations with adaptive optics, cosmological surface brightness dimming significantly affects the data, causing faint interaction signatures in the galaxy outskirts to disappear. 


Using Kinemetry, we also estimate the maximum extent of each galaxy's rotation curve. When comparing real data with mock OSIRIS observations, galaxies appear roughly 1.5 times smaller than they really are, mainly because of cosmological surface brightness dimming, which causes galaxies to lose recent interactions tracers. Although methods to increase sensitivity in outer regions - such as stacking line emissions from multiple galaxies - have been attempted, they often yield controversial results due to normalization challenges \citep{2017ApJ...840...92L, 2019MNRAS.485..934T}. For example, in the latter work, it was shown that after normalizing the rotation curves at $\sim$ 1.8 effective radius, the fraction of dark matter found is consistent with those expected from hydrodynamical simulations based on the $\Lambda$CDM theory. Furthermore, with enhanced asymmetry in galaxy's external parts, we conclude that co-adding data to improve the signal-to-noise ratio may not recover the true rotation curve shape, as signals in these regions are often random.

The mocked SINFONI observations, though able to capture fainter photons in the outer galaxy regions, suffered from low resolution, causing a loss of detailed clump structures, as well as tails and perturbations in the velocity field caused by companions and recent interactions. This blurring effect makes the galaxy appear highly symmetrical, with the rotation curve resembling a well-behaved disk. 

The combined effects of cosmic surface brightness dimming and low resolution compromise accurate estimation of the gravitational potential using only the simple kinematics of the ionized gas. In particular, low-quality velocity fields can lead to misleading interpretations of a galaxy’s dynamical state, especially if three-dimensional dynamical modeling is not employed. Consequently, relying solely on the rotation curve at large z is insufficient to accurately estimate the dark matter content of galaxies at cosmic noon.

\section{SUMMARY:}

In this paper, we investigated the influence of cosmological surface brightness dimming and loss of resolution on rotation curves extracted from galaxies in cosmic noon. For that, we used a sample of Lyman Break Analogs in z $\sim$ 0.2 and mocked observations to z $\sim$ 2.2 simulating two instruments: OSIRIS assisted with AO and SINFONI using natural seeing. 
Our main conclusions are as follows:

$\bullet$ The rotation curves extracted from mocked observations with OSIRIS are shorter than original observations; this is related to the fact that the loss of signal in the outskirts of the galaxy is exacerbated by the AO Strehl system.

$\bullet$ The rotation curves extracted from mocked observations with SINFONI benefit from better sensitivity and extend out to galaxies outskirts. However, the observations suffer from loss of resolution, resulting in smoother rotation curves.

$\bullet$ Using kinemetry, we constructed rotation curves for our real and mocked observations. We find that asymmetry in the rotation curve increases towards the galaxy outskirts, where the influence of mergers and interactions is stronger.

Cosmic surface brightness dimming, combined with low resolution, introduces significant bias in cosmic noon data, making galaxies appear smaller and more regular than they actually are. We conclude that rotation curves of distant galaxies might not be reliable enough to be used for dynamical modeling and estimating dark matter properties. The cosmic noon rotation curves might seem well behaved similar to those of cold disks, but they are such because of the resolution effects.


\section*{Acknowledgements}

We would like to thank the Hawaiian ancestry for hospitably allowing telescope operations on the summit of Mauna Kea. AEAC thanks the support of Coordination for the Improvement of Higher Education Personnel (CAPES) and CAPES PROBAL DAAD.

\section*{Data Availability}
The velocity maps and rotation curve of all LBAs are available online. Additional access will be shared on reasonable request to the corresponding author.







\begin{thebibliography}{}
\expandafter\ifx\csname natexlab\endcsname\relax\def\natexlab#1{#1}\fi
\providecommand{\url}[1]{\href{#1}{#1}}
\providecommand{\dodoi}[1]{doi:~\href{http://doi.org/#1}{\nolinkurl{#1}}}
\providecommand{\doeprint}[1]{\href{http://ascl.net/#1}{\nolinkurl{http://ascl.net/#1}}}
\providecommand{\doarXiv}[1]{\href{https://arxiv.org/abs/#1}{\nolinkurl{https://arxiv.org/abs/#1}}}

\bibitem[{{Abazajian} {et~al.}(2005){Abazajian}, {Adelman-McCarthy}, {Ag{\"u}eros}, {Allam}, {Anderson}, {Anderson}, {Annis}, {Bahcall}, {Baldry}, {Bastian}, {Berlind}, {Bernardi}, {Blanton}, {Bochanski}, {Boroski}, {Brewington}, {Briggs}, {Brinkmann}, {Brunner}, {Budav{\'a}ri}, {Carey}, {Castander}, {Connolly}, {Covey}, {Csabai}, {Dalcanton}, {Doi}, {Dong}, {Eisenstein}, {Evans}, {Fan}, {Finkbeiner}, {Friedman}, {Frieman}, {Fukugita}, {Gillespie}, {Glazebrook}, {Gray}, {Grebel}, {Gunn}, {Gurbani}, {Hall}, {Hamabe}, {Harbeck}, {Harris}, {Harris}, {Harvanek}, {Hawley}, {Hayes}, {Heckman}, {Hendry}, {Hennessy}, {Hindsley}, {Hogan}, {Hogg}, {Holmgren}, {Holtzman}, {Ichikawa}, {Ichikawa}, {Ivezi{\'c}}, {Jester}, {Johnston}, {Jorgensen}, {Juri{\'c}}, {Kent}, {Kleinman}, {Knapp}, {Kniazev}, {Kron}, {Krzesinski}, {Lamb}, {Lampeitl}, {Lee}, {Lin}, {Long}, {Loveday}, {Lupton}, {Mannery}, {Margon}, {Mart{\'\i}nez-Delgado}, {Matsubara}, {McGehee}, {McKay}, {Meiksin}, {M{\'e}nard}, {Munn}, {Nash}, {Neilsen}, {Newberg},
  {Newman}, {Nichol}, {Nicinski}, {Nieto-Santisteban}, {Nitta}, {Okamura}, {O'Mullane}, {Owen}, {Padmanabhan}, {Pauls}, {Peoples}, {Pier}, {Pope}, {Pourbaix}, {Quinn}, {Raddick}, {Richards}, {Richmond}, {Rix}, {Rockosi}, {Schlegel}, {Schneider}, {Schroeder}, {Scranton}, {Sekiguchi}, {Sheldon}, {Shimasaku}, {Silvestri}, {Smith}, {Smol{\v{c}}i{\'c}}, {Snedden}, {Stebbins}, {Stoughton}, {Strauss}, {SubbaRao}, {Szalay}, {Szapudi}, {Szkody}, {Szokoly}, {Tegmark}, {Teodoro}, {Thakar}, {Tremonti}, {Tucker}, {Uomoto}, {Vanden Berk}, {Vandenberg}, {Vogeley}, {Voges}, {Vogt}, {Walkowicz}, {Wang}, {Weinberg}, {West}, {White}, {Wilhite}, {Xu}, {Yanny}, {Yasuda}, {Yip}, {Yocum}, {York}, {Zehavi}, {Zibetti}, \& {Zucker}}]{2005AJ....129.1755A}
{Abazajian}, K., {Adelman-McCarthy}, J.~K., {Ag{\"u}eros}, M.~A., {et~al.} 2005, \aj, 129, 1755, \dodoi{10.1086/427544}

\bibitem[{{Adelberger} {et~al.}(2005){Adelberger}, {Steidel}, {Pettini}, {Shapley}, {Reddy}, \& {Erb}}]{2005ApJ...619..697A}
{Adelberger}, K.~L., {Steidel}, C.~C., {Pettini}, M., {et~al.} 2005, \apj, 619, 697, \dodoi{10.1086/426580}

\bibitem[{{Arnouts} {et~al.}(2005){Arnouts}, {Schiminovich}, {Ilbert}, {Tresse}, {Milliard}, {Treyer}, {Bardelli}, {Budavari}, {Wyder}, {Zucca}, {Le F{\`e}vre}, {Martin}, {Vettolani}, {Adami}, {Arnaboldi}, {Barlow}, {Bianchi}, {Bolzonella}, {Bottini}, {Byun}, {Cappi}, {Charlot}, {Contini}, {Donas}, {Forster}, {Foucaud}, {Franzetti}, {Friedman}, {Garilli}, {Gavignaud}, {Guzzo}, {Heckman}, {Hoopes}, {Iovino}, {Jelinsky}, {Le Brun}, {Lee}, {Maccagni}, {Madore}, {Malina}, {Marano}, {Marinoni}, {McCracken}, {Mazure}, {Meneux}, {Merighi}, {Morrissey}, {Neff}, {Paltani}, {Pell{\`o}}, {Picat}, {Pollo}, {Pozzetti}, {Radovich}, {Rich}, {Scaramella}, {Scodeggio}, {Seibert}, {Siegmund}, {Small}, {Szalay}, {Welsh}, {Xu}, {Zamorani}, \& {Zanichelli}}]{2005ApJ...619L..43A}
{Arnouts}, S., {Schiminovich}, D., {Ilbert}, O., {et~al.} 2005, \apjl, 619, L43, \dodoi{10.1086/426733}

\bibitem[{Bacon {et~al.}(2001)Bacon, Copin, Monnet, Miller, Allington-Smith, Bureau, Marcella~Carollo, Davies, Emsellem, Kuntschner, {et~al.}}]{bacon2001sauron}
Bacon, R., Copin, Y., Monnet, G., {et~al.} 2001, Monthly Notices of the Royal Astronomical Society, 326, 23

\bibitem[{{Basu-Zych} {et~al.}(2007){Basu-Zych}, {Schiminovich}, {Johnson}, {Hoopes}, {Overzier}, {Treyer}, {Heckman}, {Barlow}, {Bianchi}, {Conrow}, {Donas}, {Forster}, {Friedman}, {Lee}, {Madore}, {Martin}, {Milliard}, {Morrissey}, {Neff}, {Rich}, {Salim}, {Seibert}, {Small}, {Szalay}, {Wyder}, \& {Yi}}]{2007ApJS..173..457B}
{Basu-Zych}, A.~R., {Schiminovich}, D., {Johnson}, B.~D., {et~al.} 2007, \apjs, 173, 457, \dodoi{10.1086/521146}

\bibitem[{{Basu-Zych} {et~al.}(2009){Basu-Zych}, {Schiminovich}, {Heinis}, {Overzier}, {Heckman}, {Zamojski}, {Ilbert}, {Koekemoer}, {Barlow}, {Bianchi}, {Conrow}, {Donas}, {Forster}, {Friedman}, {Lee}, {Madore}, {Martin}, {Milliard}, {Morrissey}, {Neff}, {Rich}, {Salim}, {Seibert}, {Small}, {Szalay}, {Wyder}, \& {Yi}}]{2009ApJ...699.1307B}
{Basu-Zych}, A.~R., {Schiminovich}, D., {Heinis}, S., {et~al.} 2009, \apj, 699, 1307, \dodoi{10.1088/0004-637X/699/2/1307}

\bibitem[{Basu-Zych {et~al.}(2009)Basu-Zych, Gon{\c{c}}alves, Overzier, Law, Schiminovich, Heckman, Martin, Wyder, \& O'Dowd}]{basu2009osiris}
Basu-Zych, A.~R., Gon{\c{c}}alves, T.~S., Overzier, R., {et~al.} 2009, The Astrophysical Journal, 699, L118

\bibitem[{Binney(2005)}]{binney2005rotation}
Binney, J. 2005, Monthly Notices of the Royal Astronomical Society, 363, 937

\bibitem[{Bloom {et~al.}(2018)Bloom, Croom, Bryant, Schaefer, Bland-Hawthorn, Brough, Callingham, Cortese, Federrath, Scott, van~de Sande, D'Eugenio, Sweet, Tonini, Allen, Goodwin, Green, Konstantopoulos, Lawrence, Lorente, Medling, Owers, Richards, \& Sharp}]{10.1093/mnras/sty273}
Bloom, J.~V., Croom, S.~M., Bryant, J.~J., {et~al.} 2018, Monthly Notices of the Royal Astronomical Society, 476, 2339, \dodoi{10.1093/mnras/sty273}

\bibitem[{{Bosma}(1981{\natexlab{a}})}]{1981AJ.....86.1791B}
{Bosma}, A. 1981{\natexlab{a}}, \aj, 86, 1791, \dodoi{10.1086/113062}

\bibitem[{{Bosma}(1981{\natexlab{b}})}]{1981AJ.....86.1825B}
{Bosma},A. 1981{\natexlab{b}}, \aj, 86, 1825, \dodoi{10.1086/113063}

\bibitem[{Bouch{\'e} {et~al.}(2015)Bouch{\'e}, Carfantan, Schroetter, Michel-Dansac, \& Contini}]{bouche2015galpak3d}
Bouch{\'e}, N., Carfantan, H., Schroetter, I., Michel-Dansac, L., \& Contini, T. 2015, The Astronomical Journal, 150, 92

\bibitem[{{Bouch{\'e}} {et~al.}(2021){Bouch{\'e}}, {Bera}, {Krajnovi{\'c}}, {Emsellem}, {Mercier}, {Schaye}, {Epinat}, {Richard}, {Zoutendijk}, {Abril-Melgarejo}, {Brinchmann}, {Bacon}, {Contini}, {Boogaard}, {Wisotzki}, {Maseda}, \& {Steinmetz}}]{2021sf2a.conf..379B}
{Bouch{\'e}}, N.~F., {Bera}, S., {Krajnovi{\'c}}, D., {et~al.} 2021, in SF2A-2021: Proceedings of the Annual meeting of the French Society of Astronomy and Astrophysics, ed. A.~{Siebert}, K.~{Bailli{\'e}}, E.~{Lagadec}, N.~{Lagarde}, J.~{Malzac}, J.~B. {Marquette}, M.~{N'Diaye}, J.~{Richard}, \& O.~{Venot}, 379--382

\bibitem[{{Bundy} {et~al.}(2015){Bundy}, {Bershady}, {Law}, {Yan}, {Drory}, {MacDonald}, {Wake}, {Cherinka}, {S{\'a}nchez-Gallego}, {Weijmans}, {Thomas}, {Tremonti}, {Masters}, {Coccato}, {Diamond-Stanic}, {Arag{\'o}n-Salamanca}, {Avila-Reese}, {Badenes}, {Falc{\'o}n-Barroso}, {Belfiore}, {Bizyaev}, {Blanc}, {Bland-Hawthorn}, {Blanton}, {Brownstein}, {Byler}, {Cappellari}, {Conroy}, {Dutton}, {Emsellem}, {Etherington}, {Frinchaboy}, {Fu}, {Gunn}, {Harding}, {Johnston}, {Kauffmann}, {Kinemuchi}, {Klaene}, {Knapen}, {Leauthaud}, {Li}, {Lin}, {Maiolino}, {Malanushenko}, {Malanushenko}, {Mao}, {Maraston}, {McDermid}, {Merrifield}, {Nichol}, {Oravetz}, {Pan}, {Parejko}, {Sanchez}, {Schlegel}, {Simmons}, {Steele}, {Steinmetz}, {Thanjavur}, {Thompson}, {Tinker}, {van den Bosch}, {Westfall}, {Wilkinson}, {Wright}, {Xiao}, \& {Zhang}}]{2015ApJ...798....7B}
{Bundy}, K., {Bershady}, M.~A., {Law}, D.~R., {et~al.} 2015, \apj, 798, 7, \dodoi{10.1088/0004-637X/798/1/7}

\bibitem[{Burkert {et~al.}(2016)Burkert, Schreiber, Genzel, Lang, Tacconi, Wisnioski, Wuyts, Bandara, Beifiori, Bender, {et~al.}}]{burkert2016angular}
Burkert, A., Schreiber, N.~F., Genzel, R., {et~al.} 2016, The Astrophysical Journal, 826, 214

\bibitem[{{Cappellari} {et~al.}(2011){Cappellari}, {Emsellem}, {Krajnovi{\'c}}, {McDermid}, {Scott}, {Verdoes Kleijn}, {Young}, {Alatalo}, {Bacon}, {Blitz}, {Bois}, {Bournaud}, {Bureau}, {Davies}, {Davis}, {de Zeeuw}, {Duc}, {Khochfar}, {Kuntschner}, {Lablanche}, {Morganti}, {Naab}, {Oosterloo}, {Sarzi}, {Serra}, \& {Weijmans}}]{2011MNRAS.413..813C}
{Cappellari}, M., {Emsellem}, E., {Krajnovi{\'c}}, D., {et~al.} 2011, \mnras, 413, 813, \dodoi{10.1111/j.1365-2966.2010.18174.x}

\bibitem[{Cecil {et~al.}(2015)Cecil, Fogarty, Richards, Bland-Hawthorn, Lange, Moffett, Catinella, Cortese, Ho, Taylor, Bryant, Allen, Sweet, Croom, Driver, Goodwin, Kelvin, Green, Konstantopoulos, Owers, Lawrence, \& Lorente}]{10.1093/mnras/stv2643}
Cecil, G., Fogarty, L. M.~R., Richards, S., {et~al.} 2015, Monthly Notices of the Royal Astronomical Society, 456, 1299, \dodoi{10.1093/mnras/stv2643}

\bibitem[{Chakrabarti {et~al.}(2021)Chakrabarti, Chang, Lam, Vigeland, \& Quillen}]{chakrabarti2021measurement}
Chakrabarti, S., Chang, P., Lam, M.~T., Vigeland, S.~J., \& Quillen, A.~C. 2021, The Astrophysical Journal Letters, 907, L26

\bibitem[{Danhaive {et~al.}(2025)Danhaive, Tacchella, {\"U}bler, de~Graaff, Egami, Johnson, Sun, Arribas, Bunker, Carniani, {et~al.}}]{danhaive2025dawn}
Danhaive, A.~L., Tacchella, S., {\"U}bler, H., {et~al.} 2025, arXiv preprint arXiv:2503.21863

\bibitem[{{de Blok} {et~al.}(2008){de Blok}, {Walter}, {Brinks}, {Trachternach}, {Oh}, \& {Kennicutt}}]{2008AJ....136.2648D}
{de Blok}, W.~J.~G., {Walter}, F., {Brinks}, E., {et~al.} 2008, \aj, 136, 2648, \dodoi{10.1088/0004-6256/136/6/2648}

\bibitem[{de~Is{\'\i}dio {et~al.}(2024)de~Is{\'\i}dio, Men{\'e}ndez-Delmestre, Gon{\c{c}}alves, Grossi, Rodrigues, Garavito-Camargo, Araujo-Carvalho, Beaklini, Cavalcante-Coelho, Cortesi, {et~al.}}]{de2024dark}
de~Is{\'\i}dio, N.~G., Men{\'e}ndez-Delmestre, K., Gon{\c{c}}alves, T., {et~al.} 2024, The Astrophysical Journal, 971, 69

\bibitem[{{D'Eugenio} {et~al.}(2023){D'Eugenio}, {van der Wel}, {Piotrowska}, {Bezanson}, {Taylor}, {van de Sande}, {Baker}, {Bell}, {Bellstedt}, {Bland-Hawthorn}, {Bluck}, {Brough}, {Bryant}, {Colless}, {Cortese}, {Croom}, {Derkenne}, {van Dokkum}, {Fisher}, {Foster}, {Gallazzi}, {de Graaff}, {Groves}, {van Houdt}, {del P. Lagos}, {Looser}, {Maiolino}, {Maseda}, {Mendel}, {Nersesian}, {Pacifici}, {Poci}, {Remus}, {Sweet}, {Thater}, {Tran}, {{\"U}bler}, {Valenzuela}, {Wisnioski}, \& {Zibetti}}]{2023MNRAS.525.2789D}
{D'Eugenio}, F., {van der Wel}, A., {Piotrowska}, J.~M., {et~al.} 2023, \mnras, 525, 2789, \dodoi{10.1093/mnras/stad800}

\bibitem[{Di~Teodoro {et~al.}(2021)Di~Teodoro, Posti, Ogle, Fall, \& Jarrett}]{di2021rotation}
Di~Teodoro, E.~M., Posti, L., Ogle, P.~M., Fall, S.~M., \& Jarrett, T. 2021, Monthly Notices of the Royal Astronomical Society, 507, 5820

\bibitem[{Di~Teodoro {et~al.}(2023)Di~Teodoro, Posti, Fall, Ogle, Jarrett, Appleton, Cluver, Haynes, \& Lisenfeld}]{di2023dark}
Di~Teodoro, E.~M., Posti, L., Fall, S.~M., {et~al.} 2023, Monthly Notices of the Royal Astronomical Society, 518, 6340

\bibitem[{Evans {et~al.}(2015)Evans, Puech, Afonso, Almaini, Amram, Aussel, Barbuy, Basden, Bastian, Battaglia, {et~al.}}]{evans2015science}
Evans, C., Puech, M., Afonso, J., {et~al.} 2015, arXiv preprint arXiv:1501.04726

\bibitem[{{Fall} \& {Efstathiou}(1980)}]{1980MNRAS.193..189F}
{Fall}, S.~M., \& {Efstathiou}, G. 1980, \mnras, 193, 189, \dodoi{10.1093/mnras/193.2.189}

\bibitem[{Feng {et~al.}(2022)Feng, Shen, Yuan, Dai, \& Masters}]{Feng_2022}
Feng, S., Shen, S.-Y., Yuan, F.-T., Dai, Y.~S., \& Masters, K.~L. 2022, The Astrophysical Journal Supplement Series, 262, 6, \dodoi{10.3847/1538-4365/ac80f2}

\bibitem[{{F{\"o}rster Schreiber} {et~al.}(2006){F{\"o}rster Schreiber}, {Genzel}, {Lehnert}, {Bouch{\'e}}, {Verma}, {Erb}, {Shapley}, {Steidel}, {Davies}, {Lutz}, {Nesvadba}, {Tacconi}, {Eisenhauer}, {Abuter}, {Gilbert}, {Gillessen}, \& {Sternberg}}]{2006ApJ...645.1062F}
{F{\"o}rster Schreiber}, N.~M., {Genzel}, R., {Lehnert}, M.~D., {et~al.} 2006, \apj, 645, 1062, \dodoi{10.1086/504403}

\bibitem[{{F{\"o}rster Schreiber} {et~al.}(2009){F{\"o}rster Schreiber}, {Genzel}, {Bouch{\'e}}, {Cresci}, {Davies}, {Buschkamp}, {Shapiro}, {Tacconi}, {Hicks}, {Genel}, {Shapley}, {Erb}, {Steidel}, {Lutz}, {Eisenhauer}, {Gillessen}, {Sternberg}, {Renzini}, {Cimatti}, {Daddi}, {Kurk}, {Lilly}, {Kong}, {Lehnert}, {Nesvadba}, {Verma}, {McCracken}, {Arimoto}, {Mignoli}, \& {Onodera}}]{2009ApJ...706.1364F}
{F{\"o}rster Schreiber}, N.~M., {Genzel}, R., {Bouch{\'e}}, N., {et~al.} 2009, \apj, 706, 1364, \dodoi{10.1088/0004-637X/706/2/1364}

\bibitem[{Franx(1988)}]{Franx1988}
Franx, M. 1988, PhD thesis, University of Leiden

\bibitem[{{Genzel} {et~al.}(2017){Genzel}, {F{\"o}rster Schreiber}, {{\"U}bler}, {Lang}, {Naab}, {Bender}, {Tacconi}, {Wisnioski}, {Wuyts}, {Alexander}, {Beifiori}, {Belli}, {Brammer}, {Burkert}, {Carollo}, {Chan}, {Davies}, {Fossati}, {Galametz}, {Genel}, {Gerhard}, {Lutz}, {Mendel}, {Momcheva}, {Nelson}, {Renzini}, {Saglia}, {Sternberg}, {Tacchella}, {Tadaki}, \& {Wilman}}]{2017Natur.543..397G}
{Genzel}, R., {F{\"o}rster Schreiber}, N.~M., {{\"U}bler}, H., {et~al.} 2017, \nat, 543, 397, \dodoi{10.1038/nature21685}

\bibitem[{{Genzel} {et~al.}(2020){Genzel}, {Price}, {{\"U}bler}, {F{\"o}rster Schreiber}, {Shimizu}, {Tacconi}, {Bender}, {Burkert}, {Contursi}, {Coogan}, {Davies}, {Davies}, {Dekel}, {Herrera-Camus}, {Lee}, {Lutz}, {Naab}, {Neri}, {Nestor}, {Renzini}, {Saglia}, {Schuster}, {Sternberg}, {Wisnioski}, \& {Wuyts}}]{2020ApJ...902...98G}
{Genzel}, R., {Price}, S.~H., {{\"U}bler}, H., {et~al.} 2020, \apj, 902, 98, \dodoi{10.3847/1538-4357/abb0ea}

\bibitem[{Genzel {et~al.}(2020)Genzel, Price, Übler, Schreiber, Shimizu, Tacconi, Bender, Burkert, Contursi, Coogan, Davies, Davies, Dekel, Herrera-Camus, Lee, Lutz, Naab, Neri, Nestor, Renzini, Saglia, Schuster, Sternberg, Wisnioski, \& Wuyts}]{Genzel_2020}
Genzel, R., Price, S.~H., Übler, H., {et~al.} 2020, The Astrophysical Journal, 902, 98, \dodoi{10.3847/1538-4357/abb0ea}

\bibitem[{{Genzel} {et~al.}(2023){Genzel}, {Jolly}, {Liu}, {Price}, {Lee}, {F{\"o}rster Schreiber}, {Tacconi}, {Herrera-Camus}, {Barfety}, {Burkert}, {Cao}, {Davies}, {Dekel}, {Lee}, {Lutz}, {Naab}, {Neri}, {Nestor Shachar}, {Pastras}, {Pulsoni}, {Renzini}, {Schuster}, {Shimizu}, {Stanley}, {Sternberg}, \& {{\"U}bler}}]{2023ApJ...957...48G}
{Genzel}, R., {Jolly}, J.~B., {Liu}, D., {et~al.} 2023, \apj, 957, 48, \dodoi{10.3847/1538-4357/acef1a}

\bibitem[{{Gon{\c{c}}alves} {et~al.}(2010){Gon{\c{c}}alves}, {Basu-Zych}, {Overzier}, {Martin}, {Law}, {Schiminovich}, {Wyder}, {Mallery}, {Rich}, \& {Heckman}}]{2010ApJ...724.1373G}
{Gon{\c{c}}alves}, T.~S., {Basu-Zych}, A., {Overzier}, R., {et~al.} 2010, \apj, 724, 1373, \dodoi{10.1088/0004-637X/724/2/1373}

\bibitem[{{Gu{\'e}rou} {et~al.}(2017){Gu{\'e}rou}, {Krajnovi{\'c}}, {Epinat}, {Contini}, {Emsellem}, {Bouch{\'e}}, {Bacon}, {Michel-Dansac}, {Richard}, {Weilbacher}, {Schaye}, {Marino}, {den Brok}, \& {Erroz-Ferrer}}]{2017A&A...608A...5G}
{Gu{\'e}rou}, A., {Krajnovi{\'c}}, D., {Epinat}, B., {et~al.} 2017, \aap, 608, A5, \dodoi{10.1051/0004-6361/201730905}

\bibitem[{Heckman {et~al.}(2005)Heckman, Hoopes, Seibert, Martin, Salim, Rich, Kauffmann, Charlot, Barlow, Bianchi, Byun, Donas, Forster, Friedman, Jelinsky, Lee, Madore, Malina, Milliard, Morrissey, Neff, Schiminovich, Siegmund, Small, Szalay, Welsh, \& Wyder}]{Heckman_2005}
Heckman, T.~M., Hoopes, C.~G., Seibert, M., {et~al.} 2005, The Astrophysical Journal, 619, L35, \dodoi{10.1086/425979}

\bibitem[{Hoopes {et~al.}(2007)Hoopes, Heckman, Salim, Seibert, Tremonti, Schiminovich, Rich, Martin, Charlot, Kauffmann, {et~al.}}]{hoopes2007diverse}
Hoopes, C.~G., Heckman, T.~M., Salim, S., {et~al.} 2007, The Astrophysical Journal Supplement Series, 173, 441

\bibitem[{Jones {et~al.}(2021)Jones, Vergani, Romano, Ginolfi, Fudamoto, Béthermin, Fujimoto, Lemaux, Morselli, Capak, Cassata, Faisst, Le Fèvre, Schaerer, Silverman, Yan, Boquien, Cimatti, Dessauges-Zavadsky, Ibar, Maiolino, Rizzo, Talia, \& Zamorani}]{10.1093/mnras/stab2226}
Jones, G.~C., Vergani, D., Romano, M., {et~al.} 2021, Monthly Notices of the Royal Astronomical Society, 507, 3540, \dodoi{10.1093/mnras/stab2226}

\bibitem[{Katz {et~al.}(2016)Katz, Lelli, McGaugh, Di~Cintio, Brook, \& Schombert}]{10.1093/mnras/stw3101}
Katz, H., Lelli, F., McGaugh, S.~S., {et~al.} 2016, Monthly Notices of the Royal Astronomical Society, 466, 1648, \dodoi{10.1093/mnras/stw3101}

\bibitem[{Krajnovic {et~al.}(2006)Krajnovic, Cappellari, De~Zeeuw, \& Copin}]{krajnovic2006kinemetry}
Krajnovic, D., Cappellari, M., De~Zeeuw, P.~T., \& Copin, Y. 2006, Monthly Notices of the Royal Astronomical Society, 366, 787

\bibitem[{{Lang} {et~al.}(2017){Lang}, {F{\"o}rster Schreiber}, {Genzel}, {Wuyts}, {Wisnioski}, {Beifiori}, {Belli}, {Bender}, {Brammer}, {Burkert}, {Chan}, {Davies}, {Fossati}, {Galametz}, {Kulkarni}, {Lutz}, {Mendel}, {Momcheva}, {Naab}, {Nelson}, {Saglia}, {Seitz}, {Tacchella}, {Tacconi}, {Tadaki}, {{\"U}bler}, {van Dokkum}, \& {Wilman}}]{2017ApJ...840...92L}
{Lang}, P., {F{\"o}rster Schreiber}, N.~M., {Genzel}, R., {et~al.} 2017, \apj, 840, 92, \dodoi{10.3847/1538-4357/aa6d82}

\bibitem[{{Larkin} {et~al.}(2006){Larkin}, {Barczys}, {Krabbe}, {Adkins}, {Aliado}, {Amico}, {Brims}, {Campbell}, {Canfield}, {Gasaway}, {Honey}, {Iserlohe}, {Johnson}, {Kress}, {LaFreniere}, {Lyke}, {Magnone}, {Magnone}, {McElwain}, {Moon}, {Quirrenbach}, {Skulason}, {Song}, {Spencer}, {Weiss}, \& {Wright}}]{2006SPIE.6269E..1AL}
{Larkin}, J., {Barczys}, M., {Krabbe}, A., {et~al.} 2006, in Society of Photo-Optical Instrumentation Engineers (SPIE) Conference Series, Vol. 6269, Ground-based and Airborne Instrumentation for Astronomy, ed. I.~S. {McLean} \& M.~{Iye}, 62691A, \dodoi{10.1117/12.672061}

\bibitem[{Law {et~al.}(2006)Law, Steidel, \& Erb}]{law2006predictions}
Law, D.~R., Steidel, C.~C., \& Erb, D.~K. 2006, The Astronomical Journal, 131, 70

\bibitem[{Law {et~al.}(2009)Law, Steidel, Erb, Larkin, Pettini, Shapley, \& Wright}]{Law_2009}
Law, D.~R., Steidel, C.~C., Erb, D.~K., {et~al.} 2009, The Astrophysical Journal, 697, 2057, \dodoi{10.1088/0004-637X/697/2/2057}

\bibitem[{{Lelli} {et~al.}(2023){Lelli}, {Zhang}, {Bisbas}, {Lin}, {Papadopoulos}, {Schombert}, {Di Teodoro}, {Marasco}, \& {McGaugh}}]{2023A&A...672A.106L}
{Lelli}, F., {Zhang}, Z.-Y., {Bisbas}, T.~G., {et~al.} 2023, \aap, 672, A106, \dodoi{10.1051/0004-6361/202245105}

\bibitem[{Li {et~al.}(2020)Li, Lelli, McGaugh, \& Schombert}]{li2020comprehensive}
Li, P., Lelli, F., McGaugh, S., \& Schombert, J. 2020, The Astrophysical Journal Supplement Series, 247, 31

\bibitem[{Mancera~Pi{\~n}a {et~al.}(2022)Mancera~Pi{\~n}a, Fraternali, Oosterloo, Adams, Teodoro, Bacchini, \& Iorio}]{mancera2022impact}
Mancera~Pi{\~n}a, P.~E., Fraternali, F., Oosterloo, T., {et~al.} 2022, Monthly Notices of the Royal Astronomical Society, 514, 3329

\bibitem[{Mancera~Pi{\~n}a {et~al.}(2021{\natexlab{a}})Mancera~Pi{\~n}a, Posti, Fraternali, Adams, \& Oosterloo}]{pina2021baryonic}
Mancera~Pi{\~n}a, P.~E., Posti, L., Fraternali, F., Adams, E.~A., \& Oosterloo, T. 2021{\natexlab{a}}, Astronomy \& Astrophysics, 647, A76

\bibitem[{Mancera~Pi{\~n}a {et~al.}(2021{\natexlab{b}})Mancera~Pi{\~n}a, Posti, Pezzulli, Fraternali, Fall, Oosterloo, \& Adams}]{pina2021tight}
Mancera~Pi{\~n}a, P.~E., Posti, L., Pezzulli, G., {et~al.} 2021{\natexlab{b}}, Astronomy \& Astrophysics, 651, L15

\bibitem[{Mancera~Pi{\~n}a {et~al.}(2025)Mancera~Pi{\~n}a, Read, Kim, Marasco, Benavides, Glowacki, Pezzulli, \& Lagos}]{mancera2025galaxy}
Mancera~Pi{\~n}a, P.~E., Read, J.~I., Kim, S., {et~al.} 2025, arXiv e-prints, arXiv

\bibitem[{Marasco {et~al.}(2020)Marasco, Posti, Oman, Famaey, Cresci, \& Fraternali}]{marasco2020massive}
Marasco, A., Posti, L., Oman, K., {et~al.} 2020, Astronomy \& Astrophysics, 640, A70

\bibitem[{McGaugh {et~al.}(2001)McGaugh, Rubin, \& De~Blok}]{mcgaugh2001high}
McGaugh, S.~S., Rubin, V.~C., \& De~Blok, W. 2001, The Astronomical Journal, 122, 2381

\bibitem[{{Mistele} {et~al.}(2024){Mistele}, {McGaugh}, {Lelli}, {Schombert}, \& {Li}}]{2024ApJ...969L...3M}
{Mistele}, T., {McGaugh}, S., {Lelli}, F., {Schombert}, J., \& {Li}, P. 2024, \apjl, 969, L3, \dodoi{10.3847/2041-8213/ad54b0}

\bibitem[{Neeleman {et~al.}(2020)Neeleman, Prochaska, Kanekar, \& Rafelski}]{neeleman2020cold}
Neeleman, M., Prochaska, J.~X., Kanekar, N., \& Rafelski, M. 2020, Nature, 581, 269

\bibitem[{Nelson {et~al.}(2019)Nelson, Springel, Pillepich, Rodriguez-Gomez, Torrey, Genel, Vogelsberger, Pakmor, Marinacci, Weinberger, {et~al.}}]{nelson2019illustristng}
Nelson, D., Springel, V., Pillepich, A., {et~al.} 2019, Computational Astrophysics and Cosmology, 6, 1

\bibitem[{Osman \& Bekki(2017)}]{osman2017strong}
Osman, O., \& Bekki, K. 2017, Monthly Notices of the Royal Astronomical Society: Letters, 471, L87

\bibitem[{Overzier {et~al.}(2010)Overzier, Heckman, Schiminovich, Basu-Zych, Gon{\c{c}}alves, Martin, \& Rich}]{overzier2010morphologies}
Overzier, R.~A., Heckman, T., Schiminovich, D., {et~al.} 2010, The Astrophysical Journal, 710, 979

\bibitem[{Overzier {et~al.}(2009)Overzier, Heckman, Tremonti, Armus, Basu-Zych, Gonçalves, Rich, Martin, Ptak, Schiminovich, Ford, Madore, \& Seibert}]{Overzier_2009}
Overzier, R.~A., Heckman, T.~M., Tremonti, C., {et~al.} 2009, The Astrophysical Journal, 706, 203, \dodoi{10.1088/0004-637X/706/1/203}

\bibitem[{{Overzier} {et~al.}(2011){Overzier}, {Heckman}, {Wang}, {Armus}, {Buat}, {Howell}, {Meurer}, {Seibert}, {Siana}, {Basu-Zych}, {Charlot}, {Gon{\c{c}}alves}, {Martin}, {Neill}, {Rich}, {Salim}, \& {Schiminovich}}]{2011ApJ...726L...7O}
{Overzier}, R.~A., {Heckman}, T.~M., {Wang}, J., {et~al.} 2011, \apjl, 726, L7, \dodoi{10.1088/2041-8205/726/1/L7}

\bibitem[{Peebles(1969)}]{peebles1969origin}
Peebles, P.~J. 1969, Astrophysical Journal, vol. 155, p. 393, 155, 393

\bibitem[{Pillepich {et~al.}(2018)Pillepich, Springel, Nelson, Genel, Naiman, Pakmor, Hernquist, Torrey, Vogelsberger, Weinberger, {et~al.}}]{pillepich2018simulating}
Pillepich, A., Springel, V., Nelson, D., {et~al.} 2018, Monthly Notices of the Royal Astronomical Society, 473, 4077

\bibitem[{Pillepich {et~al.}(2019)Pillepich, Nelson, Springel, Pakmor, Torrey, Weinberger, Vogelsberger, Marinacci, Genel, van~der Wel, {et~al.}}]{pillepich2019first}
Pillepich, A., Nelson, D., Springel, V., {et~al.} 2019, Monthly Notices of the Royal Astronomical Society, 490, 3196

\bibitem[{{Planck Collaboration} {et~al.}(2016){Planck Collaboration}, {Ade}, {Aghanim}, {Arnaud}, {Ashdown}, {Aumont}, {Baccigalupi}, {Banday}, {Barreiro}, {Bartlett}, {Bartolo}, {Battaner}, {Battye}, {Benabed}, {Beno{\^\i}t}, {Benoit-L{\'e}vy}, {Bernard}, {Bersanelli}, {Bielewicz}, {Bock}, {Bonaldi}, {Bonavera}, {Bond}, {Borrill}, {Bouchet}, {Boulanger}, {Bucher}, {Burigana}, {Butler}, {Calabrese}, {Cardoso}, {Catalano}, {Challinor}, {Chamballu}, {Chary}, {Chiang}, {Chluba}, {Christensen}, {Church}, {Clements}, {Colombi}, {Colombo}, {Combet}, {Coulais}, {Crill}, {Curto}, {Cuttaia}, {Danese}, {Davies}, {Davis}, {de Bernardis}, {de Rosa}, {de Zotti}, {Delabrouille}, {D{\'e}sert}, {Di Valentino}, {Dickinson}, {Diego}, {Dolag}, {Dole}, {Donzelli}, {Dor{\'e}}, {Douspis}, {Ducout}, {Dunkley}, {Dupac}, {Efstathiou}, {Elsner}, {En{\ss}lin}, {Eriksen}, {Farhang}, {Fergusson}, {Finelli}, {Forni}, {Frailis}, {Fraisse}, {Franceschi}, {Frejsel}, {Galeotta}, {Galli}, {Ganga}, {Gauthier}, {Gerbino}, {Ghosh}, {Giard},
  {Giraud-H{\'e}raud}, {Giusarma}, {Gjerl{\o}w}, {Gonz{\'a}lez-Nuevo}, {G{\'o}rski}, {Gratton}, {Gregorio}, {Gruppuso}, {Gudmundsson}, {Hamann}, {Hansen}, {Hanson}, {Harrison}, {Helou}, {Henrot-Versill{\'e}}, {Hern{\'a}ndez-Monteagudo}, {Herranz}, {Hildebrandt}, {Hivon}, {Hobson}, {Holmes}, {Hornstrup}, {Hovest}, {Huang}, {Huffenberger}, {Hurier}, {Jaffe}, {Jaffe}, {Jones}, {Juvela}, {Keih{\"a}nen}, {Keskitalo}, {Kisner}, {Kneissl}, {Knoche}, {Knox}, {Kunz}, {Kurki-Suonio}, {Lagache}, {L{\"a}hteenm{\"a}ki}, {Lamarre}, {Lasenby}, {Lattanzi}, {Lawrence}, {Leahy}, {Leonardi}, {Lesgourgues}, {Levrier}, {Lewis}, {Liguori}, {Lilje}, {Linden-V{\o}rnle}, {L{\'o}pez-Caniego}, {Lubin}, {Mac{\'\i}as-P{\'e}rez}, {Maggio}, {Maino}, {Mandolesi}, {Mangilli}, {Marchini}, {Maris}, {Martin}, {Martinelli}, {Mart{\'\i}nez-Gonz{\'a}lez}, {Masi}, {Matarrese}, {McGehee}, {Meinhold}, {Melchiorri}, {Melin}, {Mendes}, {Mennella}, {Migliaccio}, {Millea}, {Mitra}, {Miville-Desch{\^e}nes}, {Moneti}, {Montier}, {Morgante}, {Mortlock},
  {Moss}, {Munshi}, {Murphy}, {Naselsky}, {Nati}, {Natoli}, {Netterfield}, {N{\o}rgaard-Nielsen}, {Noviello}, {Novikov}, {Novikov}, {Oxborrow}, {Paci}, {Pagano}, {Pajot}, {Paladini}, {Paoletti}, {Partridge}, {Pasian}, {Patanchon}, {Pearson}, {Perdereau}, {Perotto}, {Perrotta}, {Pettorino}, {Piacentini}, {Piat}, {Pierpaoli}, {Pietrobon}, {Plaszczynski}, {Pointecouteau}, {Polenta}, {Popa}, {Pratt}, {Pr{\'e}zeau}, {Prunet}, {Puget}, {Rachen}, {Reach}, {Rebolo}, {Reinecke}, {Remazeilles}, {Renault}, {Renzi}, {Ristorcelli}, {Rocha}, {Rosset}, {Rossetti}, {Roudier}, {Rouill{\'e} d'Orfeuil}, {Rowan-Robinson}, {Rubi{\~n}o-Mart{\'\i}n}, {Rusholme}, {Said}, {Salvatelli}, {Salvati}, {Sandri}, {Santos}, {Savelainen}, {Savini}, {Scott}, {Seiffert}, {Serra}, {Shellard}, {Spencer}, {Spinelli}, {Stolyarov}, {Stompor}, {Sudiwala}, {Sunyaev}, {Sutton}, {Suur-Uski}, {Sygnet}, {Tauber}, {Terenzi}, {Toffolatti}, {Tomasi}, {Tristram}, {Trombetti}, {Tucci}, {Tuovinen}, {T{\"u}rler}, {Umana}, {Valenziano}, {Valiviita}, {Van Tent},
  {Vielva}, {Villa}, {Wade}, {Wandelt}, {Wehus}, {White}, {White}, {Wilkinson}, {Yvon}, {Zacchei}, \& {Zonca}}]{2016A&A...594A..13P}
{Planck Collaboration}, {Ade}, P.~A.~R., {Aghanim}, N., {et~al.} 2016, \aap, 594, A13, \dodoi{10.1051/0004-6361/201525830}

\bibitem[{Posti {et~al.}(2019)Posti, Fraternali, \& Marasco}]{posti2019peak}
Posti, L., Fraternali, F., \& Marasco, A. 2019, Astronomy \& Astrophysics, 626, A56

\bibitem[{Price {et~al.}(2021)Price, Shimizu, Genzel, {\"U}bler, Schreiber, Tacconi, Davies, Coogan, Lutz, Wuyts, {et~al.}}]{price2021rotation}
Price, S., Shimizu, T., Genzel, R., {et~al.} 2021, The Astrophysical Journal, 922, 143

\bibitem[{{Puglisi} {et~al.}(2023){Puglisi}, {Dudzevi{\v{c}}i{\={u}}t{\.{e}}}, {Swinbank}, {Gillman}, {Tiley}, {Bower}, {Cirasuolo}, {Cortese}, {Glazebrook}, {Harrison}, {Ibar}, {Molina}, {Obreschkow}, {Oman}, {Schaller}, {Shankar}, \& {Sharples}}]{2023MNRAS.524.2814P}
{Puglisi}, A., {Dudzevi{\v{c}}i{\={u}}t{\.{e}}}, U., {Swinbank}, M., {et~al.} 2023, \mnras, 524, 2814, \dodoi{10.1093/mnras/stad1966}

\bibitem[{Read {et~al.}(2019)Read, Walker, \& Steger}]{10.1093/mnras/sty3404}
Read, J.~I., Walker, M.~G., \& Steger, P. 2019, Monthly Notices of the Royal Astronomical Society, 484, 1401, \dodoi{10.1093/mnras/sty3404}

\bibitem[{Ren {et~al.}(2019)Ren, Kwa, Kaplinghat, \& Yu}]{ren2019reconciling}
Ren, T., Kwa, A., Kaplinghat, M., \& Yu, H.-B. 2019, Physical Review X, 9, 031020

\bibitem[{{Rizzo} {et~al.}(2022){Rizzo}, {Kohandel, M.}, {Pallottini, A.}, {Zanella, A.}, {Ferrara, A.}, {Vallini, L.}, \& {Toft, S.}}]{rizzo}
{Rizzo}, {Kohandel, M.}, {Pallottini, A.}, {et~al.} 2022, A\&A, 667, A5, \dodoi{10.1051/0004-6361/202243582}

\bibitem[{{Rizzo} {et~al.}(2020){Rizzo}, {Vegetti}, {Powell}, {Fraternali}, {McKean}, {Stacey}, \& {White}}]{2020Natur.584..201Rizzo}
{Rizzo}, {Vegetti}, S., {Powell}, D., {et~al.} 2020, \nat, 584, 201, \dodoi{10.1038/s41586-020-2572-6}

\bibitem[{Rubin \& Ford~Jr(1970)}]{rubin1970rotation}
Rubin, V.~C., \& Ford~Jr, W.~K. 1970, The Astrophysical Journal, 159, 379

\bibitem[{Schreiber {et~al.}(2009)Schreiber, Genzel, Bouché, Cresci, Davies, Buschkamp, Shapiro, Tacconi, Hicks, Genel, Shapley, Erb, Steidel, Lutz, Eisenhauer, Gillessen, Sternberg, Renzini, Cimatti, Daddi, Kurk, Lilly, Kong, Lehnert, Nesvadba, Verma, McCracken, Arimoto, Mignoli, \& Onodera}]{Forster}
Schreiber, N. M.~F., Genzel, R., Bouché, N., {et~al.} 2009, The Astrophysical Journal, 706, 1364, \dodoi{10.1088/0004-637X/706/2/1364}

\bibitem[{Shachar {et~al.}(2023)Shachar, Price, Schreiber, Genzel, Shimizu, Tacconi, {\"U}bler, Burkert, Davies, Dekel, {et~al.}}]{shachar2023rc100}
Shachar, A.~N., Price, S., Schreiber, N.~F., {et~al.} 2023, The Astrophysical Journal, 944, 78

\bibitem[{{Shapiro} {et~al.}(2008){Shapiro}, {Genzel}, {F{\"o}rster Schreiber}, {Tacconi}, {Bouch{\'e}}, {Cresci}, {Davies}, {Eisenhauer}, {Johansson}, {Krajnovi{\'c}}, {Lutz}, {Naab}, {Arimoto}, {Arribas}, {Cimatti}, {Colina}, {Daddi}, {Daigle}, {Erb}, {Hernandez}, {Kong}, {Mignoli}, {Onodera}, {Renzini}, {Shapley}, \& {Steidel}}]{2008ApJ...682..231S}
{Shapiro}, K.~L., {Genzel}, R., {F{\"o}rster Schreiber}, N.~M., {et~al.} 2008, \apj, 682, 231, \dodoi{10.1086/587133}

\bibitem[{Snyder {et~al.}(2015)Snyder, Torrey, Lotz, Genel, McBride, Vogelsberger, Pillepich, Nelson, Sales, Sijacki, {et~al.}}]{snyder2015galaxy}
Snyder, G.~F., Torrey, P., Lotz, J.~M., {et~al.} 2015, Monthly Notices of the Royal Astronomical Society, 454, 1886

\bibitem[{{Stevens} {et~al.}(2016){Stevens}, {Croton}, \& {Mutch}}]{2016MNRAS.461..859S}
{Stevens}, A. R.~H., {Croton}, D.~J., \& {Mutch}, S.~J. 2016, \mnras, 461, 859, \dodoi{10.1093/mnras/stw1332}

\bibitem[{Swaters {et~al.}(2009)Swaters, Sancisi, Van~Albada, \& Van Der~Hulst}]{swaters2009rotation}
Swaters, R., Sancisi, R., Van~Albada, T., \& Van Der~Hulst, J. 2009, Astronomy \& Astrophysics, 493, 871

\bibitem[{Swaters {et~al.}(2000)Swaters, Madore, \& Trewhella}]{Swaters_2000}
Swaters, R.~A., Madore, B.~F., \& Trewhella, M. 2000, The Astrophysical Journal, 531, L107, \dodoi{10.1086/312540}

\bibitem[{Swinbank {et~al.}(2017)Swinbank, Harrison, Trayford, Schaller, Smail, Schaye, Theuns, Smit, Alexander, Bacon, Bower, Contini, Crain, de~Breuck, Decarli, Epinat, Fumagalli, Furlong, Galametz, Johnson, Lagos, Richard, Vernet, Sharples, Sobral, \& Stott}]{10.1093/mnras/stx201}
Swinbank, A.~M., Harrison, C.~M., Trayford, J., {et~al.} 2017, Monthly Notices of the Royal Astronomical Society, 467, 3140, \dodoi{10.1093/mnras/stx201}

\bibitem[{Teodoro \& Fraternali(2015)}]{teodoro20153d}
Teodoro, E.~D., \& Fraternali, F. 2015, Monthly Notices of the Royal Astronomical Society, 451, 3021

\bibitem[{Tiley(2020)}]{tiley2020galaxy}
Tiley, A. 2020, Galaxy disk observed to have formed shortly after the Big Bang,  Nature Publishing Group UK London

\bibitem[{{Tiley} {et~al.}(2019){Tiley}, {Swinbank}, {Harrison}, {Smail}, {Turner}, {Schaller}, {Stott}, {Sobral}, {Theuns}, {Sharples}, {Gillman}, {Bower}, {Bunker}, {Best}, {Richard}, {Bacon}, {Bureau}, {Cirasuolo}, \& {Magdis}}]{2019MNRAS.485..934T}
{Tiley}, A.~L., {Swinbank}, A.~M., {Harrison}, C.~M., {et~al.} 2019, \mnras, 485, 934, \dodoi{10.1093/mnras/stz428}

\bibitem[{Tim~de Zeeuw {et~al.}(2002)Tim~de Zeeuw, Bureau, Emsellem, Bacon, Marcella~Carollo, Copin, Davies, Kuntschner, Miller, Monnet, {et~al.}}]{tim2002sauron}
Tim~de Zeeuw, P., Bureau, M., Emsellem, E., {et~al.} 2002, Monthly Notices of the Royal Astronomical Society, 329, 513

\bibitem[{{\"U}bler {et~al.}(2021){\"U}bler, Genel, Sternberg, Genzel, Price, F{\"o}rster~Schreiber, Shimizu, Pillepich, Nelson, Burkert, {et~al.}}]{ubler2021kinematics}
{\"U}bler, H., Genel, S., Sternberg, A., {et~al.} 2021, Monthly Notices of the Royal Astronomical Society, 500, 4597

\bibitem[{Varidel {et~al.}(2019)Varidel, Croom, Lewis, Brewer, Di~Teodoro, Bland-Hawthorn, Bryant, Federrath, Foster, Glazebrook, {et~al.}}]{varidel2019sami}
Varidel, M.~R., Croom, S.~M., Lewis, G.~F., {et~al.} 2019, Monthly Notices of the Royal Astronomical Society, 485, 4024

\bibitem[{Vogelsberger {et~al.}(2014)Vogelsberger, Genel, Springel, Torrey, Sijacki, Xu, Snyder, Nelson, \& Hernquist}]{vogelsberger2014introducing}
Vogelsberger, M., Genel, S., Springel, V., {et~al.} 2014, Monthly Notices of the Royal Astronomical Society, 444, 1518

\bibitem[{{Vogelsberger} {et~al.}(2014){Vogelsberger}, {Genel}, {Springel}, {Torrey}, {Sijacki}, {Xu}, {Snyder}, {Bird}, {Nelson}, \& {Hernquist}}]{2014Natur.509..177V}
{Vogelsberger}, M., {Genel}, S., {Springel}, V., {et~al.} 2014, \nat, 509, 177, \dodoi{10.1038/nature13316}

\bibitem[{Vogelsberger {et~al.}(2014)Vogelsberger, Genel, Springel, Torrey, Sijacki, Xu, Snyder, Nelson, \& Hernquist}]{10.1093/mnras/stu1536}
Vogelsberger, M., Genel, S., Springel, V., {et~al.} 2014, Monthly Notices of the Royal Astronomical Society, 444, 1518, \dodoi{10.1093/mnras/stu1536}

\bibitem[{{White}(1984)}]{1984ApJ...286...38W}
{White}, S.~D.~M. 1984, \apj, 286, 38, \dodoi{10.1086/162573}

\bibitem[{Wisnioski {et~al.}(2015)Wisnioski, Schreiber, Wuyts, Wuyts, Bandara, Wilman, Genzel, Bender, Davies, Fossati, {et~al.}}]{wisnioski2015kmos3d}
Wisnioski, E., Schreiber, N.~F., Wuyts, S., {et~al.} 2015, The Astrophysical Journal, 799, 209

\bibitem[{{Wisnioski} {et~al.}(2019){Wisnioski}, {F{\"o}rster Schreiber}, {Fossati}, {Mendel}, {Wilman}, {Genzel}, {Bender}, {Wuyts}, {Davies}, {{\"U}bler}, {Bandara}, {Beifiori}, {Belli}, {Brammer}, {Chan}, {Davies}, {Fabricius}, {Galametz}, {Lang}, {Lutz}, {Nelson}, {Momcheva}, {Price}, {Rosario}, {Saglia}, {Seitz}, {Shimizu}, {Tacconi}, {Tadaki}, {van Dokkum}, \& {Wuyts}}]{2019ApJ...886..124W}
{Wisnioski}, E., {F{\"o}rster Schreiber}, N.~M., {Fossati}, M., {et~al.} 2019, \apj, 886, 124, \dodoi{10.3847/1538-4357/ab4db8}

\bibitem[{{Xiao} {et~al.}(2022){Xiao}, {Wang}, {Elbaz}, {Iono}, {Lu}, {Bing}, {Daddi}, {Magnelli}, {G{\'o}mez-Guijarro}, {Bournaud}, {Gu}, {Jin}, {Valentino}, {Zanella}, {Gobat}, {Martin}, {Brammer}, {Kohno}, {Schreiber}, {Ciesla}, {Yu}, \& {Okumura}}]{2022A&A...664A..63X}
{Xiao}, M.~Y., {Wang}, T., {Elbaz}, D., {et~al.} 2022, \aap, 664, A63, \dodoi{10.1051/0004-6361/202142843}

\bibitem[{Zavala \& Frenk(2019)}]{zavala2019dark}
Zavala, J., \& Frenk, C.~S. 2019, Galaxies, 7, 81

\bibitem[{Zhu {et~al.}(2017)Zhu, van~den Bosch, van~de Ven, Lyubenova, Falcón-Barroso, Meidt, Martig, Shen, Li, Yildirim, Walcher, \& Sanchez}]{10.1093/mnras/stx2409}
Zhu, L., van~den Bosch, R., van~de Ven, G., {et~al.} 2017, Monthly Notices of the Royal Astronomical Society, 473, 3000, \dodoi{10.1093/mnras/stx2409}

\end{thebibliography}
\bibliographystyle{aasjournal}

\appendix
\renewcommand{\thefigure}{A.\arabic{figure}}
\setcounter{figure}{0}
\section{Additional Rotation Curves}

\begin{figure}[ht!]
    \centering
    \includegraphics[width=\linewidth]{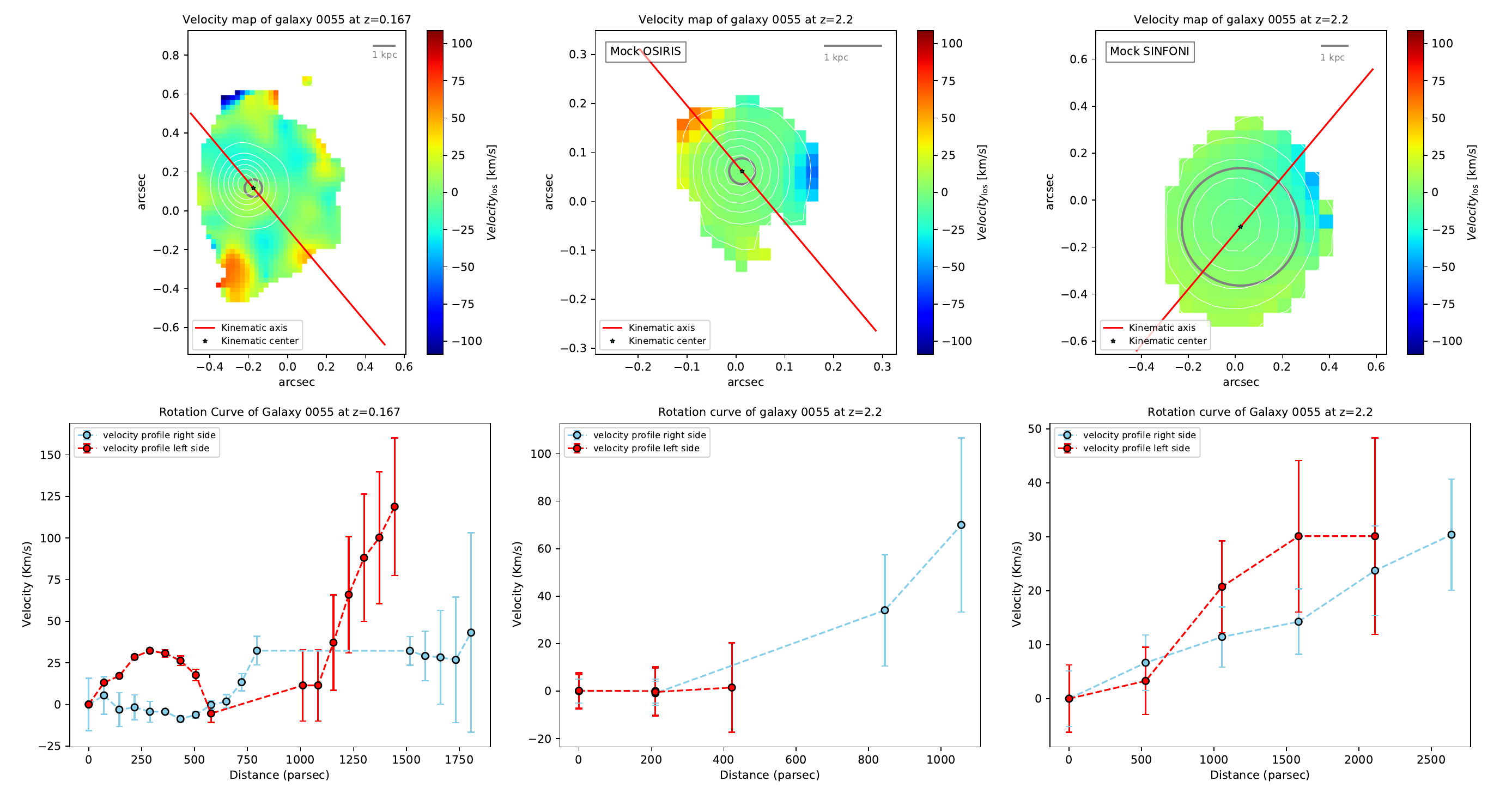}
    \caption{From left to right velocity map of the galaxy extracted from (original) Pa-$\alpha$ observations using OSIRIS instrument assisted by AO, followed by velocity map extracted for mocked OSIRIS H$\alpha$ observations at z$\sim$2.2, assisted by AO and lastly velocity map extracted for mocked SINFONI H$\alpha$ observations at z$\sim$2.2, without AO. The axes display the angular scale in arcseconds, maintaining a consistent orientation across all panels, with north pointing up and east to the left. Each panel shows the FWHM of a point source, represented by a grey circle centered on the galaxy, as an indicator of spatial resolution, along with a bar indicating the physical scale corresponding to 1 kpc at the galaxy’s redshift. The red line represents the kinematic axis used to extract the rotation curves. White contours represent the signal-to-noise distribution of the emission line map. Bottom: Rotation curve corrected for inclination extracted along the kinematic axis for original Pa-$\alpha$ observations(left), followed by mocked OSIRIS observation at z $\sim$ 2.2 (middle) and lastly mocked SINFONI observation at z$\sim$ 2.2 (right).}
    \label{rotcurvesapp}
\end{figure}

\begin{figure}[ht!]
    \centering
    \includegraphics[width=\linewidth]{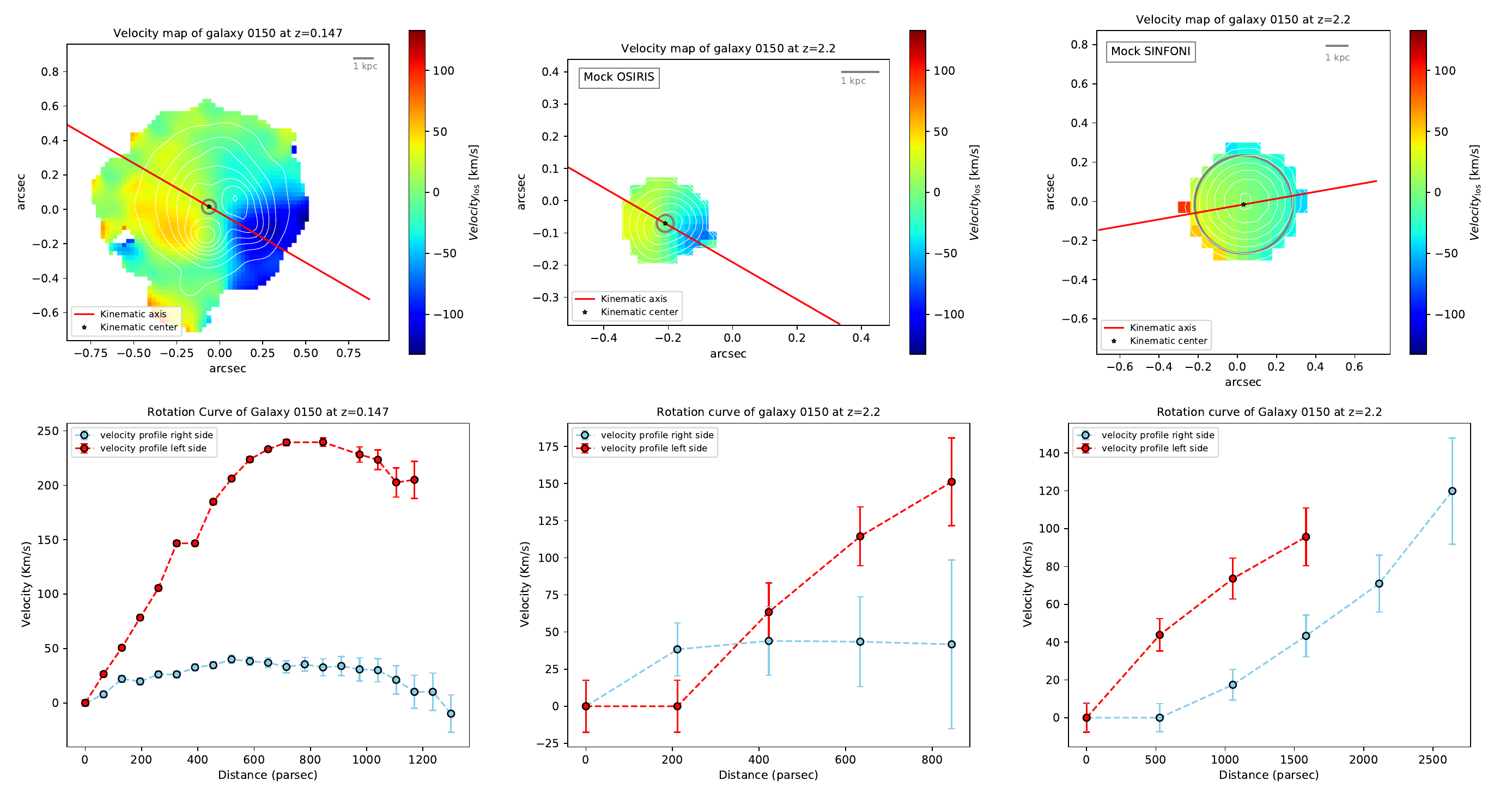}
    \vspace{5mm}
    \hrule
    \vspace{10mm}
    \includegraphics[width=\linewidth]{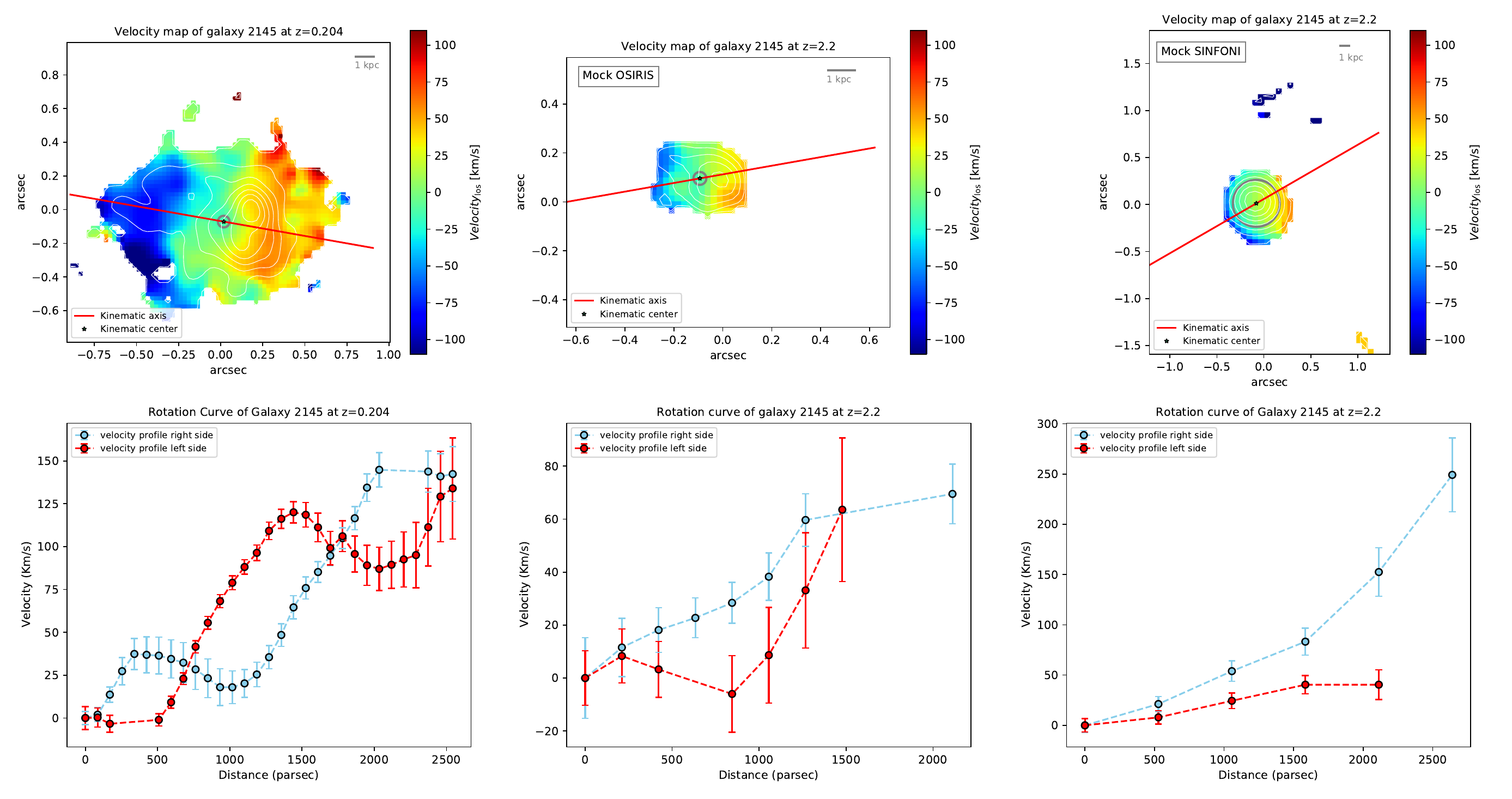}
    \caption{Same as Figure~\ref{rotcurvesapp}.}
    \label{lastrcs}
\end{figure}

From Figure \ref{rotcurvesapp} to \ref{lastrcs}, we present the velocity maps and corresponding rotation curves for the full original sample of LBAs, along with their mock observations. 

\begin{figure}[ht!]
    \centering
    \includegraphics[width=\linewidth]{rc2.pdf}
    \vspace{5mm}
    \hrule
    \vspace{10mm}
    \includegraphics[width=\linewidth]{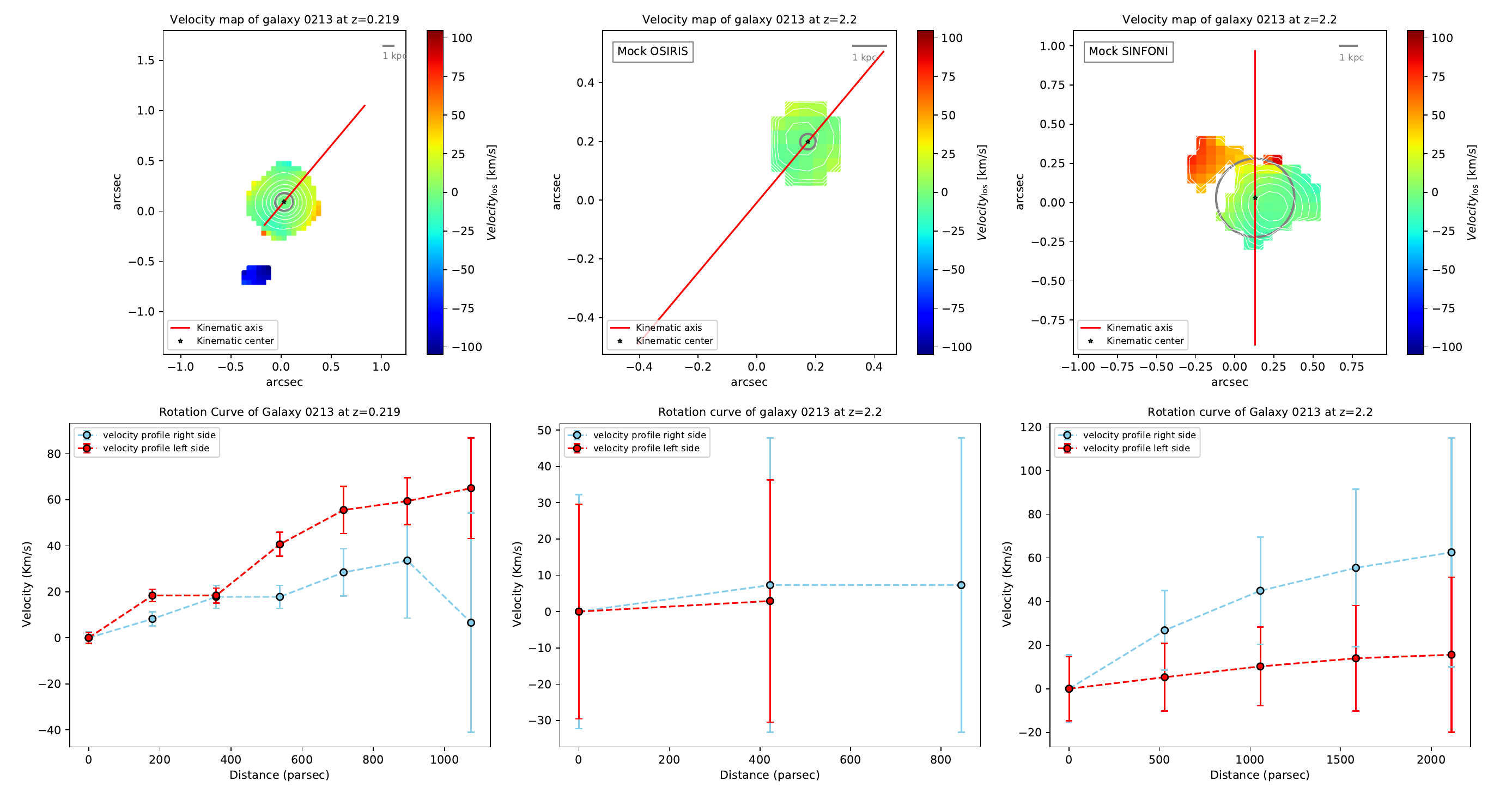}
    \caption{Same as Figure~\ref{rotcurvesapp}.}
\end{figure}

\begin{figure}[ht!]
    \centering
    \includegraphics[width=\linewidth]{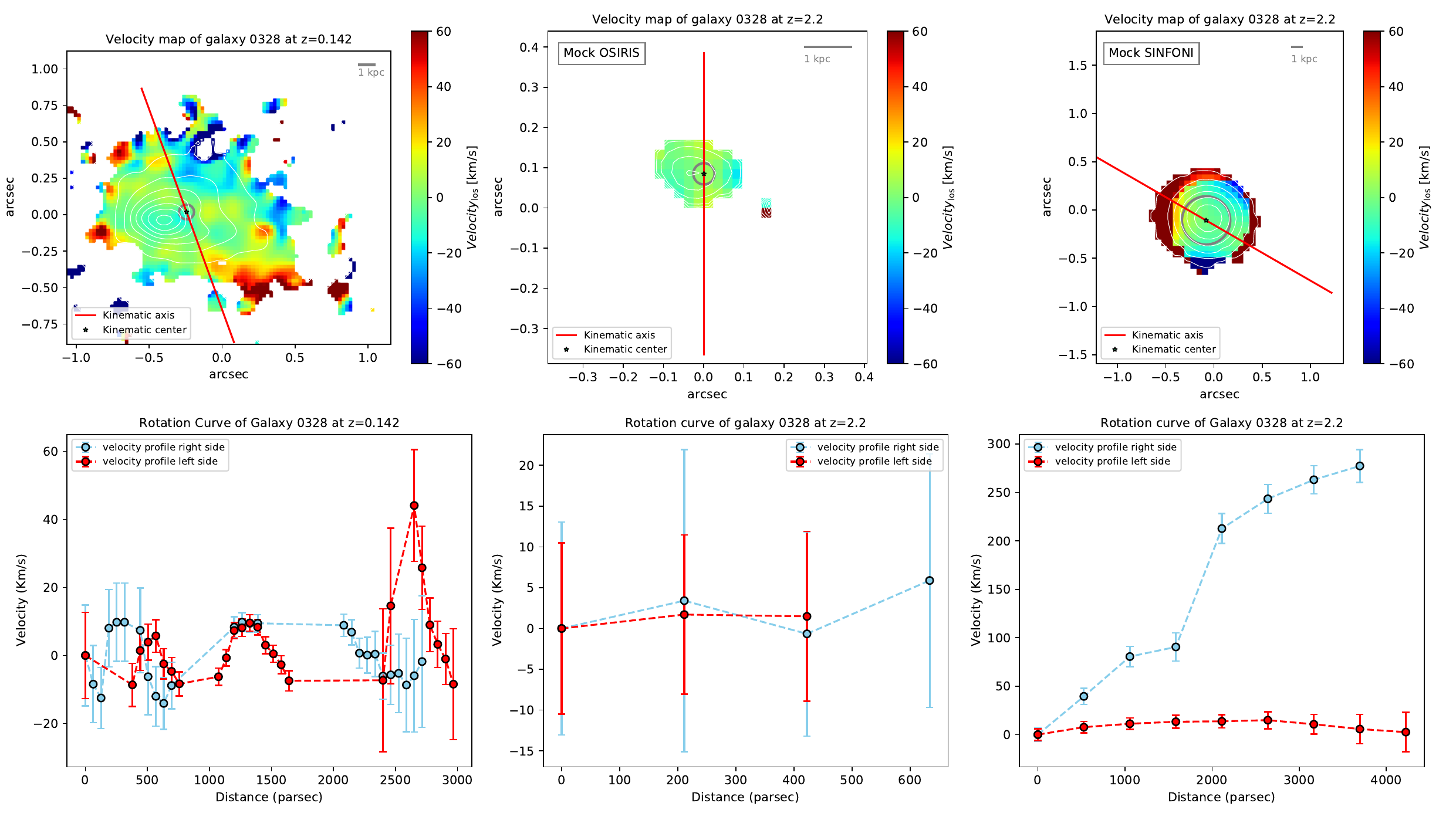}
    \vspace{5mm}
    \hrule
    \vspace{10mm}
    \includegraphics[width=\linewidth]{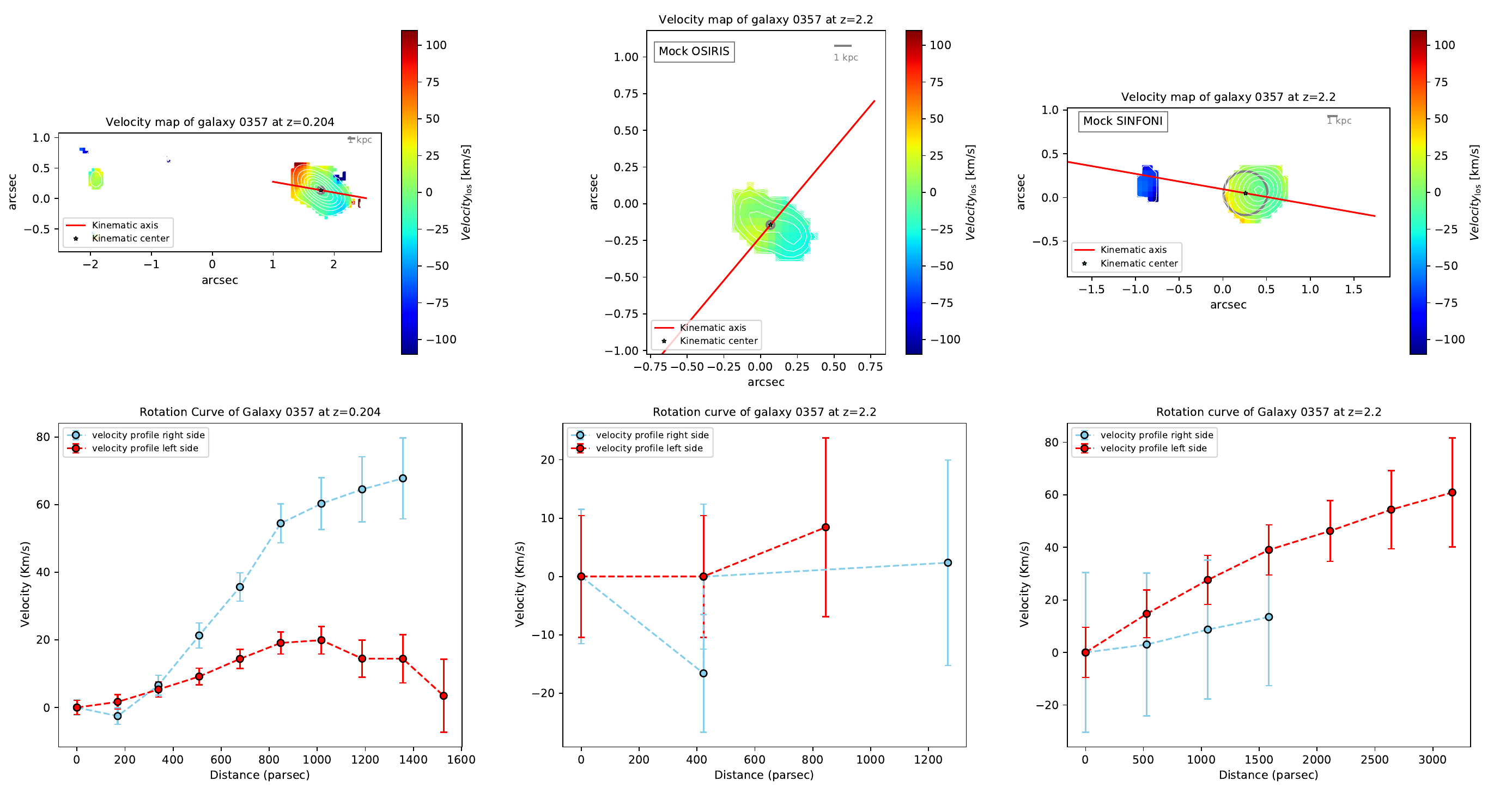}
    \caption{Same as Figure~\ref{rotcurvesapp}.}
\end{figure}

\begin{figure}[ht!]
    \centering
    \includegraphics[width=\linewidth]{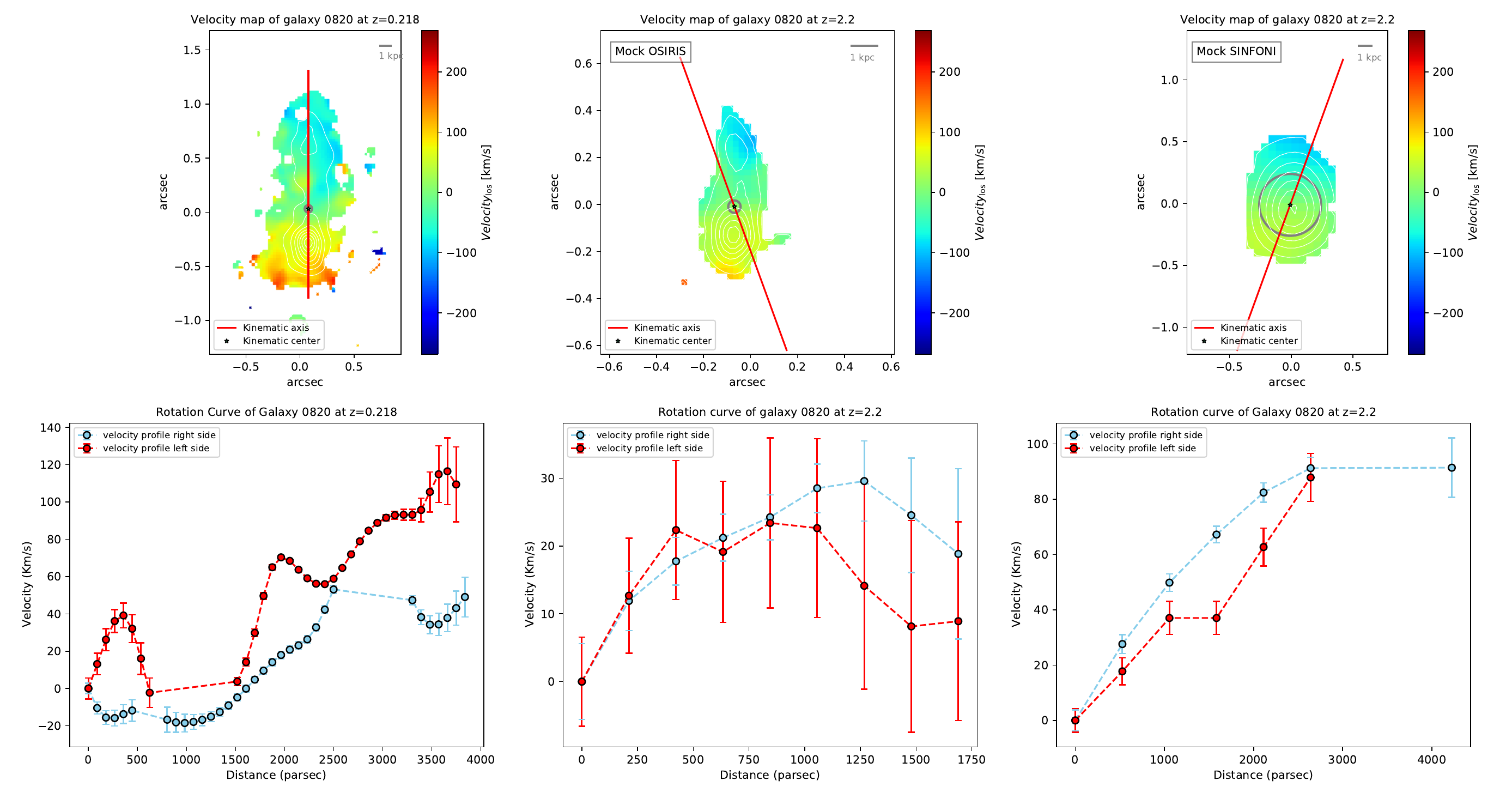}
    \vspace{5mm}
    \hrule
    \vspace{10mm}
    \includegraphics[width=\linewidth]{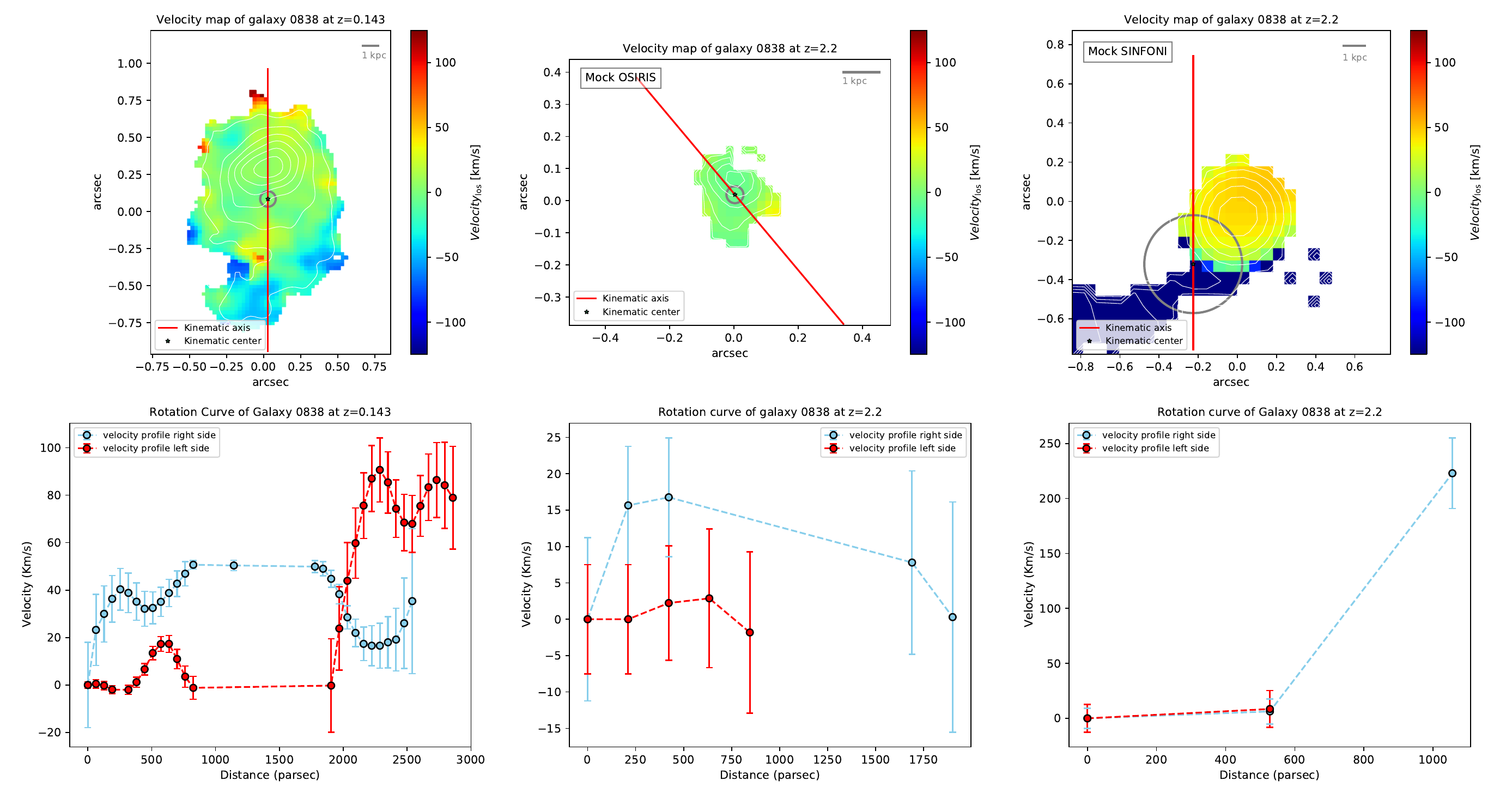}
    \caption{Same as Figure~\ref{rotcurvesapp}.}
\end{figure}

\begin{figure}[ht!]
    \centering
    \includegraphics[width=\linewidth]{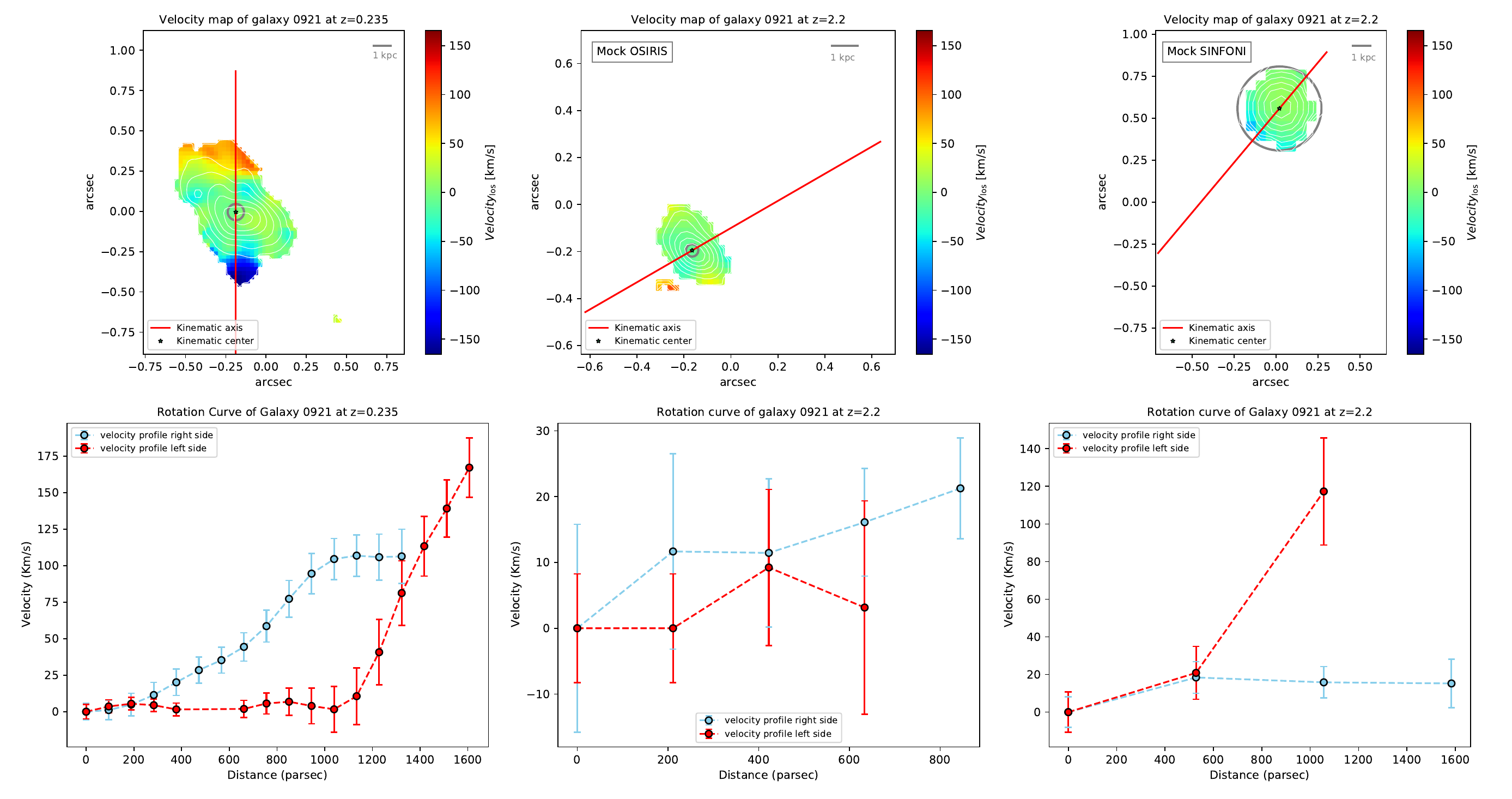}
    \vspace{5mm}
    \hrule
    \vspace{10mm}
    \includegraphics[width=\linewidth]{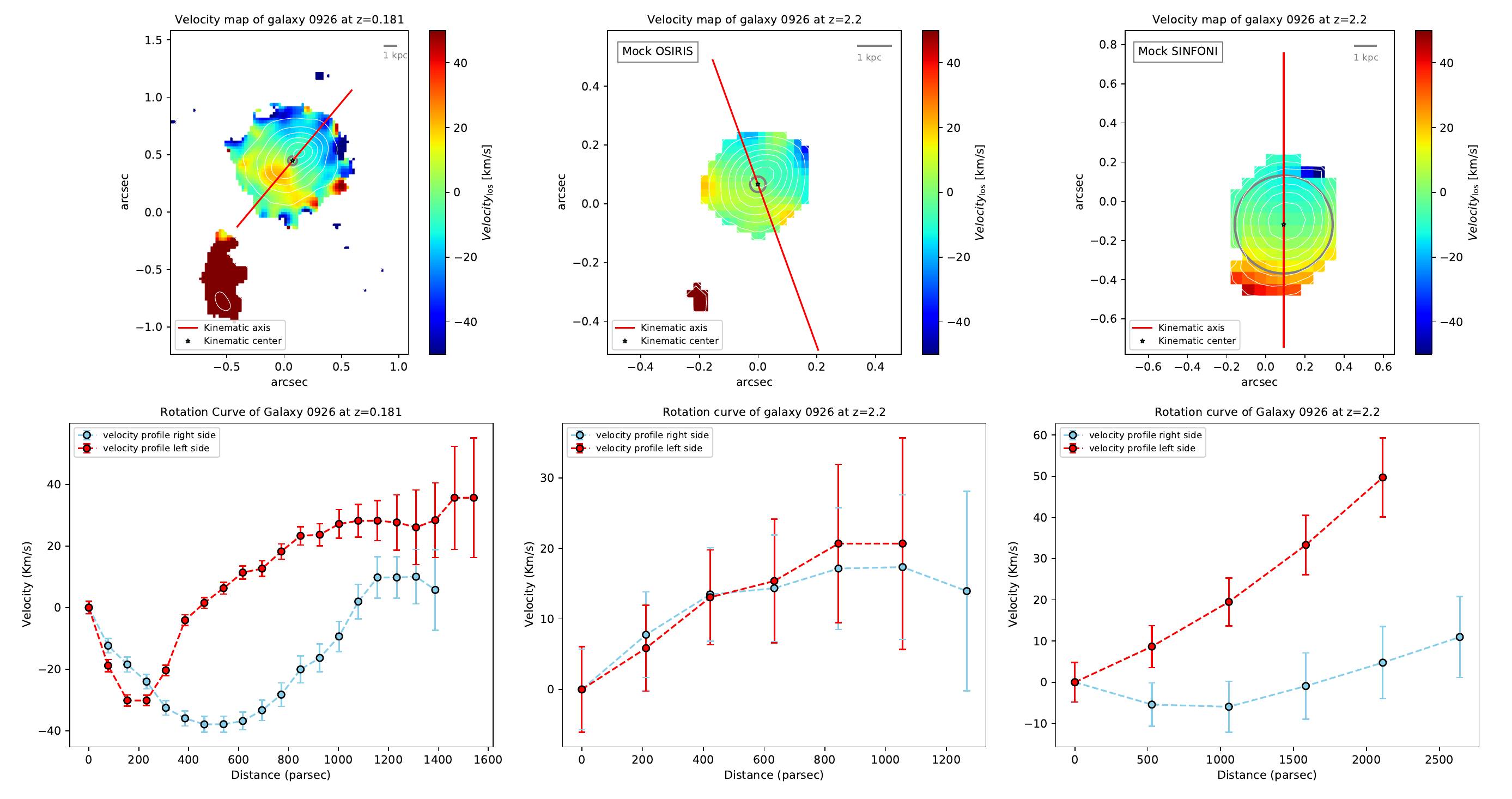}
    \caption{Same as Figure~\ref{rotcurvesapp}.}
\end{figure}

\begin{figure}[ht!]
    \centering
    \includegraphics[width=\linewidth]{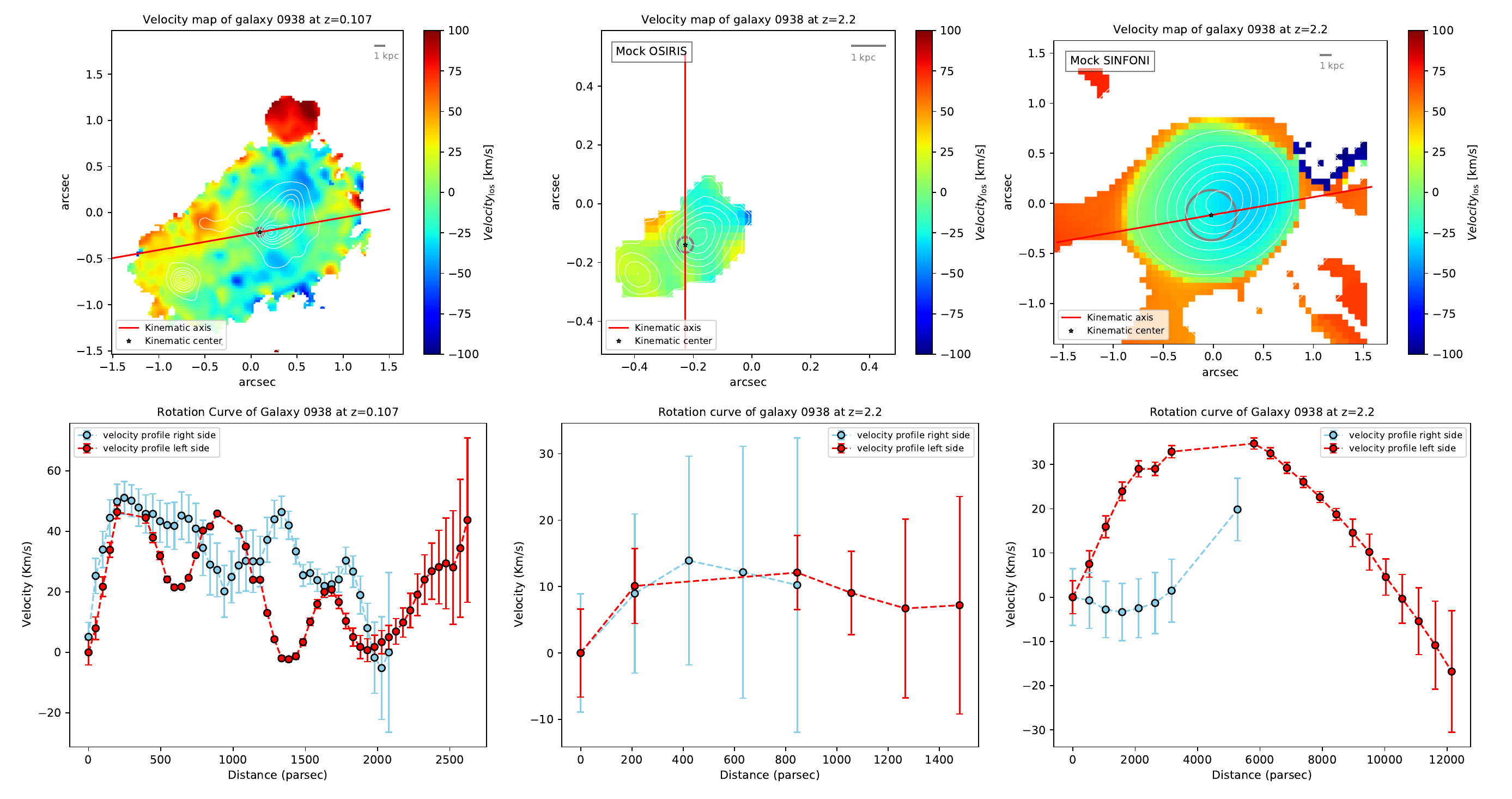}
    \vspace{5mm}
    \hrule
    \vspace{10mm}
    \includegraphics[width=\linewidth]{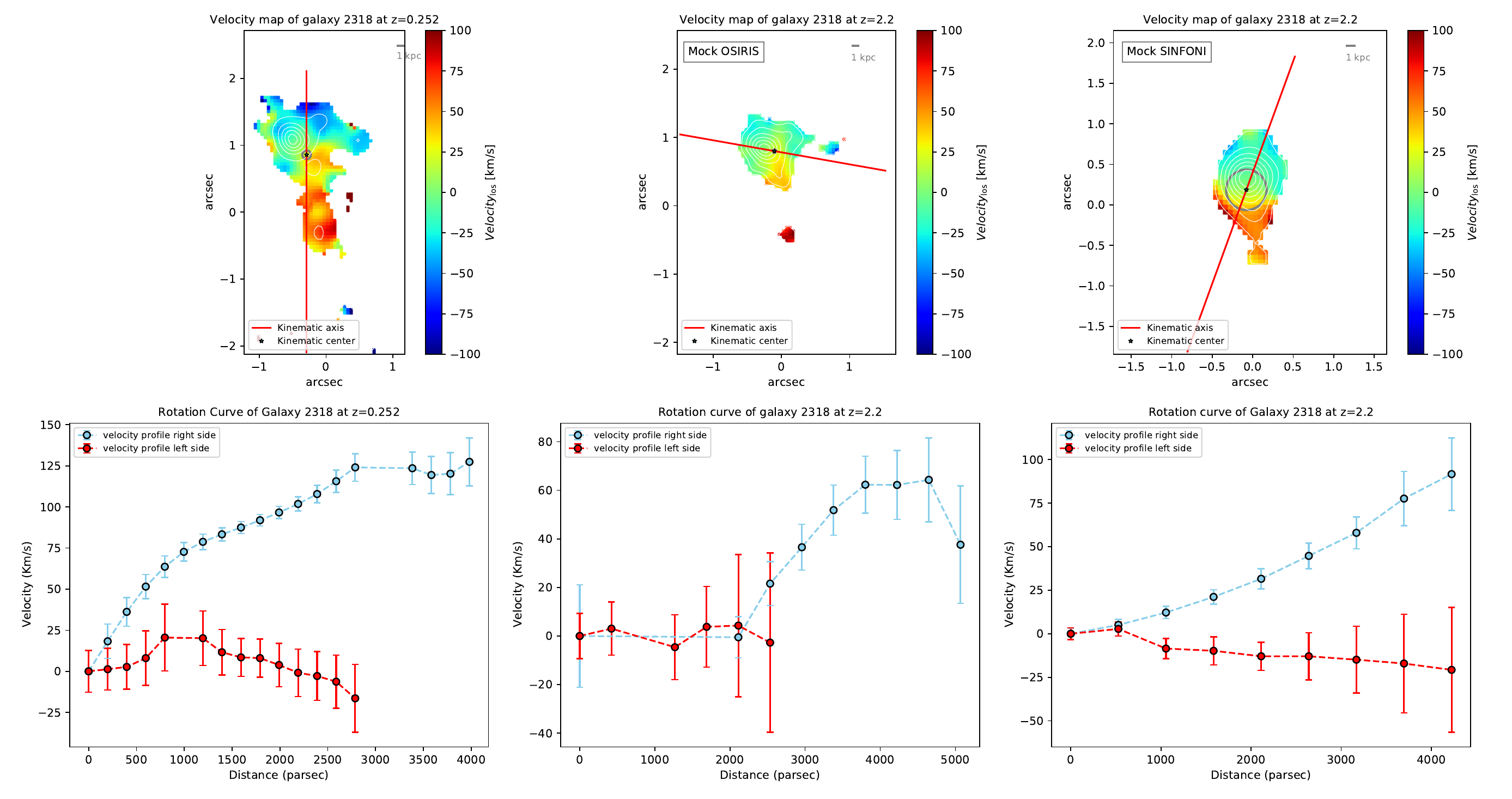}
    \caption{Same as Figure~\ref{rotcurvesapp}.}
\end{figure}

\begin{figure}[ht!]
    \centering
    \includegraphics[width=\linewidth]{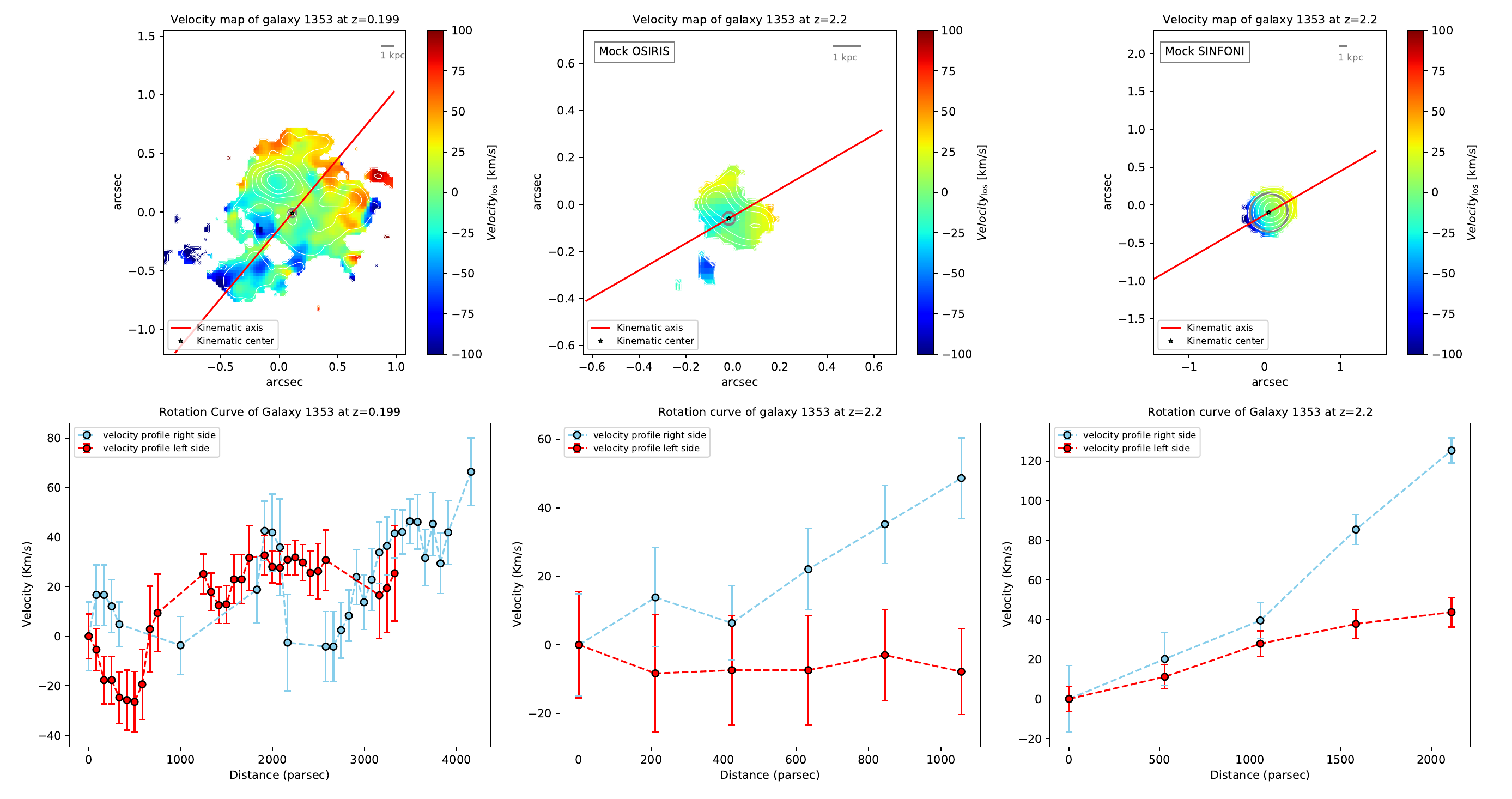}
    \vspace{5mm}
    \hrule
    \vspace{10mm}
    \includegraphics[width=\linewidth]{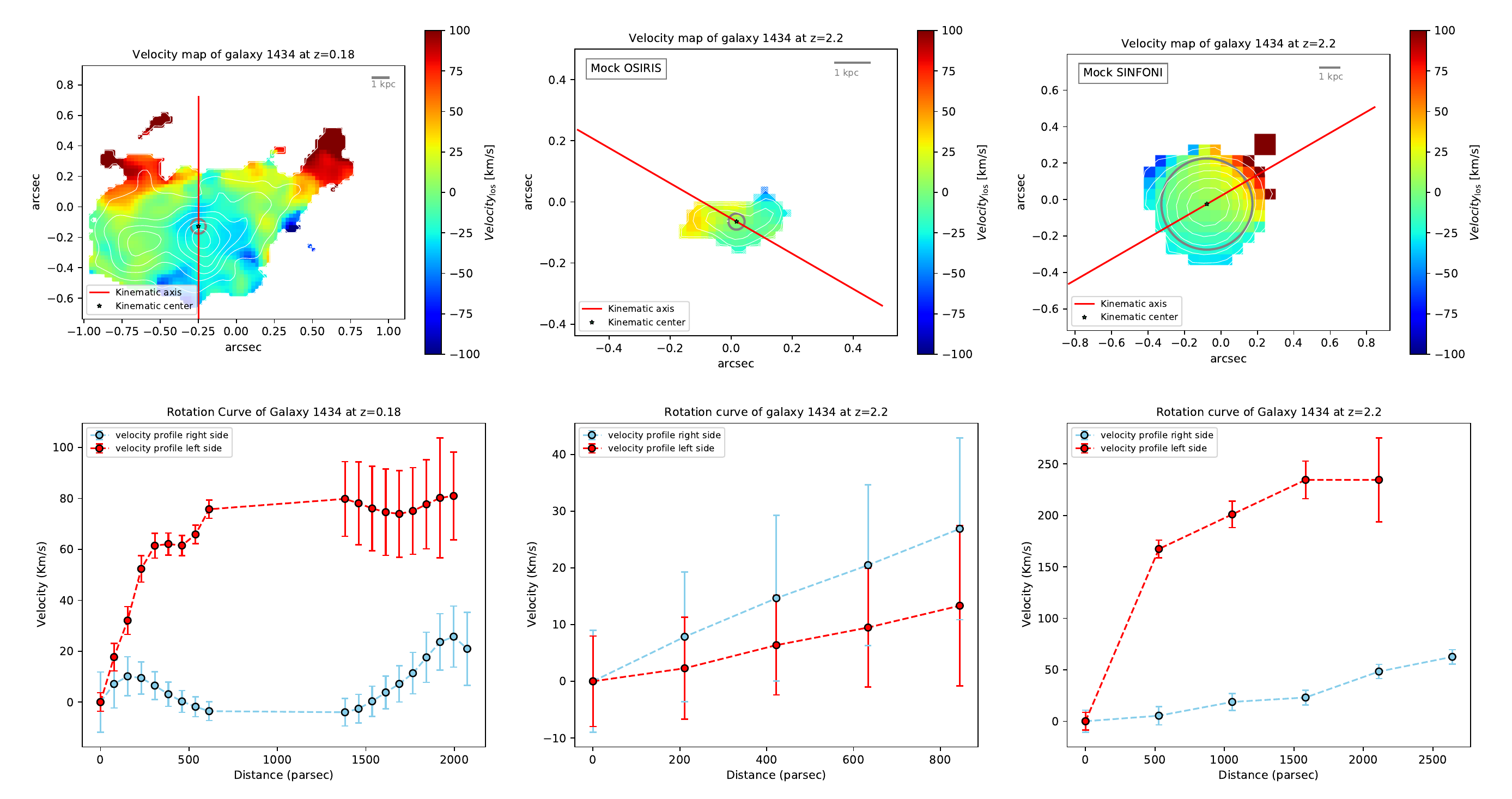}
    \caption{Same as Figure~\ref{rotcurvesapp}.}
\end{figure}

\end{document}